\documentclass{article}
\pdfoutput=1

\usepackage[a4paper, total={6in, 8in}]{geometry}
\usepackage{amsmath}
\usepackage[nohyperlinks, nolist]{acronym}
\usepackage{hyperref}
\usepackage{xspace}
\usepackage{siunitx}
\usepackage{booktabs}
\usepackage{graphicx}
\usepackage{cleveref}
\usepackage{tabularx}
\usepackage{caption}
\usepackage{subcaption}

\newcommand*{\eg}{e.g.\@\xspace}
\newcommand*{\ie}{i.e.\@\xspace}
\newcommand*{\con}{\,|\,}
\newcommand*{\smart}{BFT-SMaRt\xspace}
\newcommand*{\pbft}{\ac{pbft}\xspace}
\newcommand*{\iid}{i.i.d.\xspace}

\DeclareMathOperator\erf{erf}
\graphicspath{{../images/}{images/}{images/eval/}}

\begin{acronym}
	\acro{pdf}[PDF]{probability density function}
	\acro{smr}[SMR]{state machine replication}
	\acro{srn}[SRN]{Stochastic Reward Net}
	\acro{bft}[BFT]{Byzantine fault tolerant}
	\acro{pbft}[PBFT]{Practical Byzantine Fault Tolerance}
	\acro{ecc}[ECC]{error correction code}
\end{acronym}

\title{Bernoulli Meets PBFT: Modeling BFT Protocols in the Presence of Dynamic Failures}

\author{
	Martin Nischwitz\\
	Physikalisch-Technische Bundesanstalt\\
	Berlin, Germany\\
	martin.nischwitz@ptb.de
	\and
	Marko Esche\\
	Physikalisch-Technische Bundesanstalt\\
	Berlin, Germany\\
	marko.esche@ptb.de\\
	\and
	Florian Tschorsch\\
	Technische Universität Berlin\\
	Berlin, Germany\\
	florian.tschorsch@tu-berlin.de\\
}

\begin{document}

\maketitle

\begin{abstract}
The publication of the pivotal state machine replication protocol PBFT
laid the foundation for a large body of BFT protocols.
While many successors to PBFT have been developed,
there is no general technique to compare these protocols
under realistic network conditions such as unreliable links.
In this paper, we introduce a probabilistic model for evaluating BFT protocols
in the presence of dynamic link and crash failures.
Based on modeling techniques from communication theory,
the network state of replicas is captured and used to derive the success probability of the protocol execution.
To this end, we examine the influence of link and crash failure rates as well as the number of replicas.
The model is derived from the communication pattern,
making it implementation-independent
and facilitating an adaptation to other BFT protocols.
The model is validated with a simulation of PBFT and \smart.
Further, a comparison in protocol behavior of PBFT, Zyzzyva and SBFT is performed and critical failure thresholds are identified.
\end{abstract}

\section{Introduction}
\label{sec.intro}
The rapidly increasing connectivity of devices,
as for example envisioned by the Internet of things,
entices the development of large-scale, globally distributed systems.
The European Metrology Cloud project~\cite{Thiel2017a},
which aims to coordinate the digital transformation of legal metrology,
is a prime example.
The scale and complexity of such systems leads to a higher risk
of failure and/or malicious behavior.
The demand for trust and reliability, however, remains unchanged.

A technique to offer higher fault tolerance and availability for distributed systems is \ac{smr}.
It requires processes to find agreement on the order of state transitions
and, thus, consensus on the system state.
One of the most prominent \ac{bft} protocols is \pbft~\cite{Castro2002}.
Many modern systems, including the recent surge of blockchain applications,
utilize \pbft, or a variation of it, as their core consensus algorithm,
\eg, \smart~\cite{Bessani2014}, Tendermint~\cite{Buchman2018},
RBFT~\cite{Aublin2013}, CheapBFT~\cite{Kapitza2012},
and Hyperledger Fabric v0.6~\cite{Androulaki2018}.

While many advanced \ac{bft} protocols exist,
the performance impact of dynamic failures in general
and unreliable links in particular is often ignored.
Many protocols, however, require that messages arrive within a defined timespan,
\ie, there is a bound on the message delay.
If that bound is not met, the performance of the protocols might deteriorate.
To quantify performance, protocols are often compared with benchmarks on either real systems or simulations.
To the best of our knowledge, there is unfortunately no technique to assess the impact of unreliable links
on the performance of \ac{bft} protocols without requiring a comprehensive implementation.

In this paper, we fill the gap and present a probabilistic modeling approach
for \ac{bft} protocols to measure the impact of dynamic link and crash failures on their performance.
The model is derived from the communication pattern
and therefore transferable to many \ac{bft} protocols.
It predicts the system state assuming the so-called dynamic link failure model~\cite{Santoro1989},
that is, unreliable communication links with message losses and high delays.
More specifically, we assume a constant failure probability for all links and processes,
model state transitions as Bernoulli trials,
and express the resulting system state as \aclp{pdf}.
Thus, our model can provide feedback already during the design and development phase of \ac{bft} protocols
as well as support to parameterize timeouts.

Our model validation confirms that the model accurately captures the behavior of \pbft as well as \smart~\cite{Bessani2014}
and allows to predict the probability for successful protocol executions.
Moreover, we employ our model to Zyzzyva~\cite{Kotla2007}, and SBFT~\cite{GolanGueta2019} to showcase its applicability to other \ac{bft} protocols.
In our evaluation, we analyze the mentioned protocols,
most notably \pbft, with respect to the impact of various failures
and protocol stability.
Accordingly, the paper's contributions can be summarized as follows:
\begin{itemize}
	\item We develop a probabilistic model for \pbft
	that is able to quantify the impact of dynamic link and crash failures.
	\item Since the model is implementation independent,
	we are able to generalize and apply it to other \ac{bft} protocols,
	including \smart, Zyzzyva, and SBFT.
	\item We validate the approach by comparing it to a simulation of \pbft
	and \smart.
	\item We identify critical values for dynamic link failure
	and crash failure rates at which the previously mentioned protocols become unstable.
\end{itemize}

The remainder of the paper is organized as follows.
In Section~\ref{sec.related_work}, we discuss related work with a focus on \ac{bft} modeling and failures.
In Section~\ref{sec.system_model}, we define the system model.
Next, we describe the detailed derivation of our modeling approach for \pbft in Section~\ref{sec.model},
and present a simulation-based model validation in Section~\ref{sec.validation}.
In Section~\ref{sec.evaluation}, we use our model to reveal structural differences between \pbft, \smart, Zyzzyva, and SBFT,
before we conclude the paper in Section~\ref{sec.summary}.

\section{Related Work}
\label{sec.related_work}
\subsection{Preliminaries}

The main properties of \ac{bft} protocols are described by the notions of
\emph{safety} and \emph{liveness}~\cite{Castro2002}.
Safety indicates that the protocol satisfies serializability,
\ie, it behaves like a centralized system.
Liveness, on the other hand, indicates that the system will eventually respond
to requests.
In order to tolerate $f$ faulty processes,
at least $3f+1$ processes are necessary~\cite{Bracha1985}.
Aside from process-related failures,
the network, \ie, the communication between processes,
also impacts the performance of \ac{bft} protocols and is often overlooked.

The network can be described as either synchronous or asynchronous.
To bridge the gap between completely syn\-chro\-nous/asyn\-chro\-nous systems,
the term \emph{partially synchronous} was introduced~\cite{Dwork1988}.
A partially synchronous system may start in an asynchronous state but will,
after some unspecified time, eventually return to a synchronous state.
This captures temporary link failures, for example.
A different perspective on a partially synchronous system is to assume
a network with fixed upper bounds on message delays and processing times,
where both are unknown a priori.
The partially synchronous system model is utilized by many \ac{bft} protocols,
\eg, \pbft, to circumvent the FLP impossibility~\cite{Fischer1985}
and guarantee liveness during the synchronous states of the system,
without requiring it at all times.
Deterministic \ac{bft} protocols guarantee safety, even in the asynchronous state,
but require synchronous periods to guarantee liveness.

To detect Byzantine behavior,
most \ac{bft} protocols utilize timeouts (and signatures).
If the happy path of a protocol fails to make progress,
a sub-protocol, \eg, a view change protocol,
is triggered to recover~\cite{Liskov2010}.
To optimize performance, the timeout values should depend on
the bounded message delay in the synchronous periods of the network,
which plays an important role for deployments~\cite{Sa2013}.
In addition to message delay characteristics, some networks, \eg, wireless networks,
might be susceptible to link failures, leading to message omissions or corruptions.
These failures are formally captured by the so-called \emph{dynamic link failure model}~\cite{Santoro1989},
where the authors prove that consensus is impossible in a synchronous system
with an unbounded number of transmission failures.
Schmid et al.~\cite{Schmid2001,Schmid2009} introduced a hybrid failure model
to capture process and communication failures
and derived bounds on the number of failures for synchronous networks.

\subsection{BFT Models}
Other modeling techniques to analyze the performance of distributed systems
and \pbft have previously been proposed in the literature.
The framework HyPerf~\cite{Halalai2011} combines model checking
and simulation techniques to evaluate \ac{bft} protocols.
While model checking usually proves correctness,
their framework uses simulations to explore the possible paths in the model checker
and evaluate the performance of the protocol.
The model is validated against an implementation of \pbft
to predict latencies and throughput.

A method to model \pbft with \acp{srn} was proposed in~\cite{Sukhwani2017}.
The authors deployed Hyperledger Fabric v0.6, which implements \pbft,
on a cluster of four nodes and compare the collected data to the distributions of their \acp{srn}.
Afterwards, the number of nodes in the model is scaled and evaluated regarding the mean time to consensus.

Singh et al.~\cite{Singh2008} provided a more direct approach to evaluate \ac{bft} protocols
with the simulation framework BFTSim.
The framework utilizes the high-level declarative language P2~\cite{Loo2005}
to implement three different \ac{bft} protocols~\cite{Castro2002, Kotla2007, Abd-El-Malek2005}.
Built upon ns-2~\cite{ns2016}, the simulation can explore various network conditions.

While the previously listed models and/or simulators offer the possibility
to evaluate the performance of \ac{bft} protocols,
they all require \emph{comprehensive implementations} of the respective protocol.
The model presented in this paper, however, is derived from the \emph{communication pattern}.
Moreover, no simulations or measurements are required to employ our model;
all system states can be evaluated with closed-form expressions at low computational cost.
Finally, the main focus of our work is to present a model for \ac{bft} protocols
that captures the impact of unreliable communication and varying message delays,
which the other models only considered as a minor aspect.

\subsection{Link and Crash Failures}
In this paper, we analyze the consequences of message omissions in \pbft
caused by dynamic link failures, which also covers misconfigured timeouts.
Fathollahnejad et al.~\cite{Fathollahnejad2013} examined the impact of dynamic link failures
on their leader election algorithm in a traffic control system.
They assume a constant failure probability
and present techniques to calculate the probability for disagreement.
As in our paper, the number of received messages of an all-to-all broadcast is modeled
with Bernoulli trials.
Since their use case does not consider Byzantine failures,
their protocol does not require a consecutive collection of message quorums
which is, however, implemented in most \ac{bft} protocols
and thus the main focus of the model presented in this paper.

Xu et al.\ presented RATCHETA~\cite{Xu2018},
a consensus protocol which was designed for embedded devices
in a wireless network that might be prone to dynamic link failures.
They included an evaluation with artificially induced packet losses,
measuring the number of failing consensus instances.
RATCHETA requires a trusted subsystem that prevents a process
from casting differing votes during the same consensus instance,
eliminating the possibility of equivocations.
It therefore yields a $2f+1$ resilience, allowing $f$ Byzantine failures.

In addition, there is a body of literature that covers the theoretical limits of
consensus protocols~\cite{Schmid2002, Schmid2009, Biely2011},
which are based on a hybrid failure model~\cite{Schmid2001}
and therefore also capture link failures.
Existing models, however, rarely consider performance aspects
resulting from unreliable network conditions such as dynamic crash and link failures.
Since all \ac{bft} protocols have a built-in protocol to recover from crashed processes,
\eg, view changes,
their impact on the performance is tied to the frequency of the recovery algorithm execution.

\section{System Model}\label{sec.system_model}

\subsection{Process Model}\label{sec.system_model.process_model}
The distributed system consists of a fixed number of $n$ processes
(we use the term process, node, and replica interchangeably).
Typically, no more than $f$ processes are allowed to be subject to Byzantine faults and
$n \geq 3f+1$ replicas are required to guarantee safety~\cite{Bracha1985}.

To tolerate Byzantine (or crash) failures in an asynchronous setting, distributed systems rely on timeouts in combination with message thresholds to make progress. Consequently, it is common to describe the protocols in phases, \ie, system states in which each process, \eg, awaits the reception of a certain amount of messages or, alternatively, a timeout.
Our model is time-free in that all events are mapped to the respective phases of the protocol.

In order to capture diverse failure cases,
\eg, congestion due to high traffic load,
we introduce the term \emph{dynamic crash failures},
along the lines of the \emph{dynamic link failure} model by Santoro et al.~\cite{Santoro1989}.
That is, processes can become unavailable in each phase.
In this paper, dynamic crash failures are assumed to be independent and identically distributed~(\iid) random variables for all processes during each phase. A recovered process may or may not participate in later phases of the protocol, depending on the protocol.

Since our model is derived from the communication pattern of the protocol,
special roles such as the primary in \pbft which follow a different communication pattern,
are incorporated into the model.

\subsection{Network Model}\label{sec.system_model.network_model}

We assume that each network node has a peer-to-peer connection to all other nodes.
The network model in this paper allows for
(i) messages to be delayed indefinitely, \ie, past the configured timeout parameter of \pbft, and
(ii) message omissions as well as corruptions.
The former case acknowledges \ac{bft} protocols that rely on synchronous periods
to guarantee liveness and are based on timeouts to detect process and/or link failures.
\pbft, for example, makes use of timeouts to detect if progress is being made
and as a consequence to initiate the view-change protocol,
which elects a new primary.
Messages that arrive after a configured timeout can therefore be considered as message omissions.
The same applies to invalid or corrupted messages.
The resulting failure model can be described with the \emph{dynamic link failure model},
introduced in~\cite{Santoro1989}.
While in practice many \ac{bft} protocols rely on the network layer to guarantee reliable communication,
\eg, TCP, they should implement means to handle lost messages due to crashed/ma\-li\-cious processes.
We therefore assume unidirectional links, which implies unreliable communication.

If assumptions made regarding the bound of message delays fail,
\ie, the timeouts are not configured appropriately,
the protocol can be considered to operate in an asynchronous network
with unbounded message delays.
This does not apply if an attacker is considered to have control over the scheduling of messages,
as this could easily lead to stopping a \ac{bft} protocol altogether~\cite{Miller2016}.
As with process failures, the link failures are assumed to be \iid for all links.

\section{Modeling PBFT}\label{sec.model}

The model presented in this section offers means to evaluate \pbft
in the presence of dynamic link failures and crash failures.
In particular, a performance impact assessment becomes possible.
For the sake of clarity, we provide an overview of our modeling approach
and introduce our notation first.
Next, we unroll our model for various failure types step-by-step
starting with dynamic crash failures, before we incorporate dynamic link failures.

\subsection{Overview}

\begin{figure}
	\centering
	\includegraphics[width=0.7\linewidth]{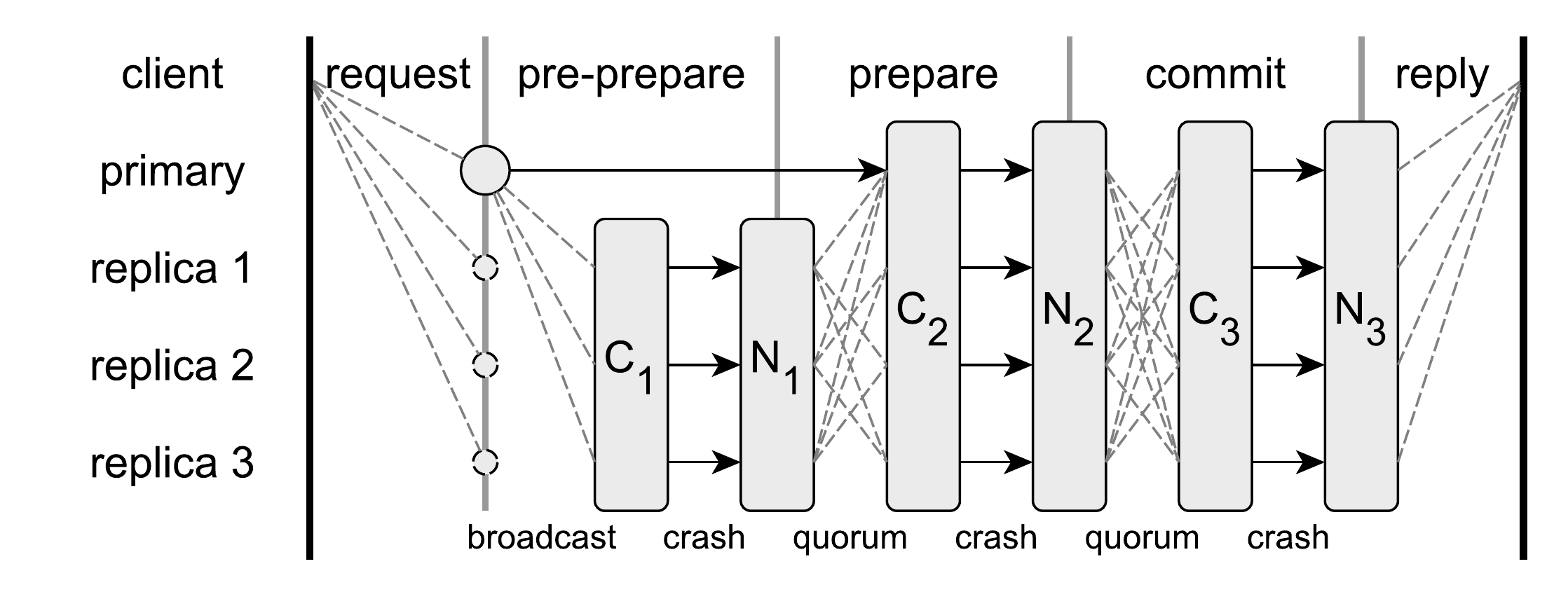}
	\caption{Modeled view on PBFT's happy path communication pattern. Each phase is modeled via alternating predictions for crash ($C_i$) and link failures ($N_i$).}
	\label{fig.pbft}
\end{figure}

In \pbft, the happy path consists of five phases of message exchanges, as depicted in \Cref{fig.pbft}.
The first and last phase consist of transmissions from and to the client.
In the first phase, the leader of the current view will collect and serialize client requests.
This is followed by a second phase in which the primary disseminates the requests to all other replicas
in a so-called \texttt{pre-prepare} message.
If a replica receives and accepts a \texttt{pre-prepare} message,
it stores that message and enters the third phase,
broadcasting \texttt{prepare} messages.
Replicas wait for a quorum of messages,
\ie, at least $2f+1$ \texttt{prepare} messages which match the stored \texttt{pre-prepare} message.
In the fourth phase replicas broadcast \texttt{commit} messages to all other replicas.
Finally, if a replica collects another quorum of \texttt{commit} messages,
which match a previously collected quorum of \texttt{prepare} messages,
it will commit (and execute) the state transition.
In the fifth and last phase of the protocol, replicas reply to the client,
confirming that the client's request was executed from the replicated system.

Disregarding the client interactions,
\pbft's happy path can be reduced to three phases
by omitting the first and last phase.
The remaining three phases can be summarized as
a broadcast phase and two quorum collection phases.
In the following, we assume that the primary is ready to initiate the consensus algorithm.
Consequently, only the communication between the replicas
is captured in our model.

In each phase of the protocol, the communication between and the availability of replicas
is modeled as a combination of Bernoulli trials.
More specifically, we model link and crash failures in alternating rounds
for each of \pbft's phases, as depicted at the bottom of \Cref{fig.pbft}.
We use random variables $N_i$ and $C_i$ to express the success probabilities
for the respective failure type in phase $i$.
We assume failures to be \iid for all links and processes.

In a first step, only faulty nodes are modeled as crash failures
in a series of interdependent Bernoulli trials,
\ie, $N_1 \rightarrow N_2 \rightarrow N_3$.
In a second step, we extend the model by incorporating link failures.
The communication is modeled along the lines of the three transmission phases
$C_1$, $C_2$, and $C_3$.
Combined with the node failures, our model yields an interleaving series of dependent system states,
\ie, $C_1 \rightarrow  N_1 \rightarrow C_2 \rightarrow  N_2 \rightarrow C_3 \rightarrow N_3$.

In summary,
the system state of all replicas at each protocol phase is captured
by a series of \acfp{pdf},
each constituting the calculation of the following.
Please note, that each \acp{pdf} allows for precise prediction of the protocol behavior and can be transformed into more common performance metrics, \eg, latency, with statistics or other models that predict the duration of individual phases.

\subsection{Notation}
In \Cref{tab.pbft}, we summarize relevant probabilities, events, and random variables,
which are used in our model.
For ease of comprehension, the link and crash failure distributions are now reduced to single probabilities, \ie, $p_l$ and $p_c$, respectively.
The assumption to have identical link failure probabilities for all links is not an uncommon practice in this field of research~\cite{Fathollahnejad2017,Singh2008,Xu2018}.
The impact of arbitrary failure distributions on the model are explained in \Cref{sec.model_discussion}.
The system state of the protocol is modeled by calculating \acp{pdf} that describe each replica's state.
To this end, the random variables and events listed in \Cref{tab.pbft}
are indexed according to \pbft's phases.
In particular, $C_1$, $C_2$, and $C_3$ represent the number of replicas that
received a \texttt{pre-prepare} message,
received a quorum of \texttt{prepare} messages and
received a quorum of \texttt{commit} messages, respectively.
Additionally, the number of active replicas after each phase is described with $N_1$, $N_2$, and $N_3$.
Due to the nature of the \pbft algorithm, the distributions are dependent on each other, \ie, a replica that crashed or failed to collect the required messages will not be able to complete the happy path.
The performance of the protocol can be assessed by calculating the previously mentioned \acp{pdf} that describe the states of each replica.

A key building block of our model are Bernoulli trials. Therefore, we use the notation
$
B(n,p,k) = \binom{n}{k} p^k (1-p)^{n-k}
$
to express the probability to get exactly $k$ successes in a Bernoulli experiment
with $n$ trials and a success probability of $p$. Furthermore, we define
$
B(n,p,[k,l]) = \sum_{i=k}^{l} B(n,p,i)
$
as the sum over all Bernoulli trials with at least $k$ and up to $l$ successes.
Finally, the notations
$P_X(x) = P(X=x)$ and $P_{X|Y}(x|y) = P(X=x|Y=y)$
are used to abbreviate (conditional) probabilities.

\begin{table}[t]
	\caption{Model notation for \pbft.}\label{tab.pbft}
	\begin{tabularx}{\columnwidth}{lXl}
		\toprule
		\textbf{Symbol} & \textbf{Description} & \textbf{Range}\\

		\midrule

		\multicolumn{3}{c}{\textsc{Probabilities}}\\
		\cmidrule(){1-3}
		\multicolumn{1}{l}{$p_c$} & Probability for a crash failure. & $[0, 1]$\\
		\multicolumn{1}{l}{$p_l$} & Probability for a link failure. & $[0, 1]$\\
		\addlinespace

		\multicolumn{3}{c}{\textsc{Events}}\\
		\cmidrule(){1-3}
		\multicolumn{1}{l}{$C_p$} & Indicates a successful reception of a quorum of \texttt{prepare} messages at the current primary.&\\
		\addlinespace

		\multicolumn{3}{c}{\textsc{Random Variables}}\\
		\cmidrule(){1-3}
		\multicolumn{1}{l}{$C_1$} & The number of replicas that have received a \texttt{pre-prepare} message. & $[0,n-1]$\\
		\multicolumn{1}{l}{$N_1$} & The number of replicas, excluding the primary, that did not crash in the \texttt{pre-prepare} phase. & $[0,n-1]$\\
		\multicolumn{1}{l}{$C_{2,n}$} & The number of replicas, excluding the primary, that have received a \texttt{pre-prepare} message as well as collected a quorum of \texttt{prepare} messages. & $[0,n-1]$\\
		\multicolumn{1}{l}{$C_2$} & The number of replicas that received a \texttt{pre-prepare} message as well as collected a quorum of \texttt{prepare} messages. & $[0,n]$\\
		\multicolumn{1}{l}{$N_2$} & The number of replicas that did not crash in the \texttt{prepare} phase. & $[0,n]$\\
		\multicolumn{1}{l}{$C_3$} & The number of replicas that have received a \texttt{pre-prepare} message as well as collected a quorum of both, \texttt{prepare} and \texttt{commit} messages. & $[0,n]$\\
		\multicolumn{1}{l}{$N_3$} & The number of replicas that have did not crash in the \texttt{commit} phase and successfully executed the algorithm. & $[0,n]$\\
		\bottomrule
	\end{tabularx}
\end{table}

\subsection{Modeling crash failures}

We start by modeling one of the most prominent failures, crash failures.
In particular,
a crash failure implies no participation of the crashed process in the phase in which the crash occurred.
Hence, the replica will neither receive nor send any messages.
In the following, the three random variables $N_1, N_2$, and $N_3$
are derived for a crash failure probability $p_c$,
assuming reliable communication links.

The happy path of \pbft is initiated with
the primary broadcasting a \texttt{pre-prepare} message to all other replicas.
For the sake of simplicity, it is assumed the primary cannot crash during the \texttt{pre-prepare} phase.
Since the probability for a crash is uniform across all nodes,
the \ac{pdf} of the still active replicas $N_1$ is given by
$P_{N_1}(n_1) = B(n,1-p_c,n_1)$.
The distribution of $N_1$ describes the number of replicas that will now broadcast \texttt{prepare} messages to all other replicas in the second phase of the protocol.

Following this procedure, the
distribution of active nodes in further phases
is calculated conditioned on the previous phase, meaning
\begin{equation}\label{eq.n2_crash}
P_{N_i}(n_i) = \sum\limits_{n_{i-1}} B(n_{i-1},1-p_c,n_i) \cdot P_{N_{i-1}}(n_{i-1}) \qquad \text{for } i=2,3.
\end{equation}
Adding up the values of $P(N_3 \geq 2f+1)$ allows to predict the success probability for the happy path of \pbft.
As replicas cannot skip a phase in \pbft,
a crashed replica will not recover
during the happy path
rendering dynamic crashes similar to permanent ones.

\subsection{Modeling crash and link failures}

We now extend our model by introducing link failures, \ie,
the links are no longer considered reliable
and are subject to a link failure probability $p_l$.
The three random variables $C_1, C_2$, and $C_3$ are introduced
to model the behavior of the protocol during the three communication phases.
Due to the special behavior of the primary in the second phase,
$C_2$ is divided into the event $C_p$ and random variable $C_{2,n}$
to capture the communication of the primary and other replicas, respectively.
Assuming the dynamic link failure model,
all links are subject to the same failure probability $p_l$
and can be, as with crash failures before, described with Bernoulli trials.
In the following, we therefore start alternating between $C_i$ and $N_i$ (cf. \Cref{fig.pbft})
to model the success of the message delivery and node availability, respectively.

\paragraph*{Calculating $C_1$}\mbox{}\\
After receiving a request from the client, the first phase is initiated with
the primary broadcasting a \texttt{pre-prepare} message to all other replicas.
Again, we assume that the primary cannot crash while receiving requests from a client.
Since the success probability for each message transmission is equal to $1-p_l$
and independent of other transmissions,
the number of successful transmissions can be calculated with a Bernoulli trial.
The \ac{pdf} of $C_1$ is given by
\begin{equation}\label{eq.c1}
P_{C_1}(c_1) = B(n-1, 1-p_l, c_1)
\end{equation}
and describes the number of replicas that have received a \texttt{pre-preapre} message
from the primary.

\paragraph*{Calculating $N_1$}\mbox{}\\
Given the distribution of $C_1$,
some of replicas might crash in this phase.
Based on $C_1$, the \ac{pdf} of the remaining active replicas $N_1$ is
\begin{equation}\label{eq.n1}
P_{N_1}(n_1) = \sum\limits_{c_1=0}^{n-1} P_{N_1|C_1}(n_1 \con c_1) \cdot P_{C_1}(c_1) = \sum\limits_{c_1=0}^{n-1} B(c_1, 1-p_c,n_1) \cdot P_{C_1}(c_1).
\end{equation}

\paragraph*{Calculating $C_2$}\mbox{}\\
The communication in the second phase is composed of two cases:
(i) whether the primary can collect $2f$ \texttt{prepare} messages (\ie, $C_p$), and
(ii) the number of non-primary replicas that collect at least $2f+1$ \texttt{prepare} messages
(\ie, $C_{2,n}$).

The primary can only collect at least $2f$ \texttt{prepare} messages
if at least $2f$ active replicas have received the previous \texttt{pre-prepare} message,
\ie, $N_1 \geq 2f$.
In this case, at least $2f$ transmissions of \texttt{prepare} messages of the $N_1$ replicas
have to successfully reach the primary.
This can be expressed as the sum over all favorable Bernoulli trials, \ie, all trials with at least $2f$ successes out of $N_1$.
The conditional probability $P(C_p \con N_1 = n_1)$ for the primary to collect the \texttt{prepare} message is given by
\begin{equation}\label{eq.c2p_conditional}
P(C_p \con N_1 = n_1) =\begin{cases}
0, & n_1 < 2f\\
B(n_1,1-p_l,[2f, n]) & \text{otherwise.}
\end{cases}
\end{equation}

For a non-primary node, \ie, a replica, to advance to $C_2$, two requirements need to be met:
(i) the replica has received a respective \texttt{pre-prepare} message, and
(ii) the replica has collected a quorum of matching \texttt{prepare} messages.
For a quorum, only $2f-1$ \texttt{prepare} messages are required,
since a replica's own \texttt{prepare} message and the primary's \texttt{pre-prepare} message
count towards the $2f+1$ required messages.
The previous requirements translate to
\begin{enumerate}
	\item there cannot be more replicas that receive $2f-1$ \texttt{prepare} messages
	than replicas that have previously received a \texttt{pre-prepare} message,
	\ie, $C_{2,n} \leq N_1$, and

	\item a replica can only receive $2f-1$ \texttt{prepare} messages
	if at least $2f$ replicas, including itself, have received a \texttt{pre-prepare},
	\ie, $N_1 \geq 2f$.
\end{enumerate}
The calculation of $C_{2,n}$ can thus be divided into the following cases,
assuming that $n_1$ replicas have received a \texttt{pre-prepare} message.
First, for $n_1 < 2f$, no replica will be able to gather the required quorum of \texttt{prepare} messages,
thus, the probability for $c_{2,n} = 0$ is always one.
Second, if $c_{2,n} > n_1$, the probability has to be zero.
Finally, for all other cases, of the $n_1$ replicas that broadcast \texttt{prepare} messages, excluding the primary,
the probability for $c_{2,n}$ replicas to receive $2f-1$ of those messages can be modeled as another Bernoulli trial.
The probability of success in that Bernoulli trial is identical to a replica receiving at least $2f-1$ messages of the $n_1 - 1$ possible.
Thus, the conditional \ac{pdf} of $C_{2,n}$ for $n_1 \geq 2f$ and $c_{2,n} \leq n_1$ is given by
\begin{equation}\label{eq.c2n_conditional}
P_{C_{2,n}|N_1}(c_{2,n} \con n_1) = B(n_1,p_2(n_1),c_{2,n})
\end{equation}
with $p_2(n_1)$ being the probability
that a replica will receive at least $2f-1$ \texttt{prepare} messages,
given that $n_1$ replicas, including the replica itself, are broadcasting that message,
which implies they have received the \texttt{pre-prepare} message as well.
This can be calculated with another Bernoulli trial to get $2f-1$ receptions from $n_1-1$ messages of the other replicas,
\begin{equation}\label{eq.c2n_conditional_helper}
p_2(n_1) = B(n_1-1,1-p_l,[2f-1, n]).
\end{equation}

Combining \eqref{eq.c2p_conditional} and \eqref{eq.c2n_conditional}
yields the conditional \ac{pdf} of $C_2$.
The calculation is split into multiple cases as follows
  \begin{equation}\label{eq.c2_conditional}
  P_{C_2|N_1}(c_2|n_1) =
  \begin{cases}
  P_{C_{2,n}|N_1}(0 \con n1) \cdot P(\overline{C_p} \con N_1=n_1), & c_{2,n} = 0\\
  P_{C_{2,n}|N_1}(n-1 \con n_1) \cdot P(C_p \con N_1=n_1), & c_{2,n} = n\\
  P_{C_{2,n}|N_1}(c_{2,n} \con n_1) \cdot P(\overline{C_p} \con N_1=n_1)\\
  \,+\,P_{C_{2,n}|N_1}(c_{2,n}-1 \con n_1) \cdot P(C_p \con N_1=n_1), & c_{2,n} \leq n_1+1\\
  0, & \text{otherwise.}
  \end{cases}
  \end{equation}

The final \ac{pdf} of $C_2$ is given by applying the law of total probability to \eqref{eq.c2_conditional},
which yields
\begin{equation}\label{eq.c2}
P_{C_2}(c_2) = \sum\limits_{n_1 = 0}^{n-1} P_{C_2|N_1}(c_2 \con n_1) \cdot P_{N_1}(n_1).
\end{equation}

\paragraph*{Calculating $N_2$}\mbox{}\\
As with $N_1$ and \eqref{eq.n1}, the distribution of replicas that are still active, based on $C_2$, is
\begin{equation}\label{eq.n2}
P_{N_2}(n_2) = \sum\limits_{c_2=0}^{n} B(c_2, 1-p_c,n_2) \cdot P_{C_2}(c_2).
\end{equation}

\paragraph*{Calculating $C_3$}\mbox{}\\
Now, let us turn to the states $C_3$ and $N_3$.
In the third phase, the primary behaves in the same way as every other replica,
simplifying many calculations regarding the communication as we do not need to mind so many exceptions.
As with $C_2$, there are two requirements necessary for a replica to reach $C_3$:
(i) the replica must be in state $C_2$ and
(ii) it must have received at least $2f$ \texttt{commit} messages, not counting its own.
Thus, we can conclude that
\begin{enumerate}
	\item there cannot be more replicas that have received $2f$ \texttt{commit} messages
	than replicas that have reached $C_2$,
	\ie, $C_3 \leq C_2$, and
	\item a replica can only receive $2f$ \texttt{commit} messages
	if at least $2f+1$ replicas, including itself, have reached state $C_2$,
	\ie, $C_2 > 2f$.
\end{enumerate}
Deriving $C_3$ is similar to $C_2$
with \eqref{eq.c2n_conditional}, \eqref{eq.c2n_conditional_helper},
and \eqref{eq.c2_conditional}.
The conditional probability of $C_3$ for $c_2 > 2f$ and $c_3 \leq c_2$ is accordingly
\begin{equation}\label{eq.c3_conditional}
P_{C_3|C_2}(c_3 \con c_2) = B(c_2,p_3(c_2),c_3)
\end{equation}
where $p_3(c_2)$ is the probability
that a replica will receive at least $2f$ \texttt{commit} messages
if $c_2$ replicas, including itself, are broadcasting that message, \ie,
\begin{equation}\label{eq.c3_helper}
p_3(c_2) = B(c_2-1,1-p_l,[2f, n]).
\end{equation}
Applying the law of total probability yields
\begin{equation}\label{eq.c3}
P_{C_3}(c_3) = \sum\limits_{c_2 = 0}^{n} P_{C_3|C_2}(c_3 \con c_2) \cdot P_{C_2}(c_2).
\end{equation}

\paragraph*{Calculating $N_3$}\mbox{}\\
Finally, the \ac{pdf} of all active nodes after the last phase is similar to \eqref{eq.n1} and \eqref{eq.n2}
\begin{equation}\label{eq.n3}
P_{N_3}(n_3) = \sum\limits_{c_3=0}^{n} B(c_3, 1-p_c,n_3) \cdot P_{C_3}(c_3).
\end{equation}

Combining \eqref{eq.c1}, \eqref{eq.n1}, \eqref{eq.c2}, \eqref{eq.n2}, \eqref{eq.c3} and \eqref{eq.n3}
yields the system state at any time during execution of \pbft's happy path.
If more than $2f$ replicas have completed the last phase,
\ie $P(N_3 > 2f)$, the happy path of \pbft was successful.
For the system to provide liveness in regards to the current request,
only $f+1$ replicas are sufficient.

\subsection{Generalization}\label{sec.model_discussion}

For the sake of simplicity, we so far assumed constant failure probabilities
for links and processes, \ie, $p_l$ and $p_c$.
We also assumed in our calculations, that those probabilities be constant for each phase of \pbft.
This is not a requirement and could be expanded to reflect more sophisticated failure models
that include time-based correlations, for example, as long as they remain \iid for each phase.

Since the model is derived solely from communication patterns,
it can be adapted to other \ac{bft} protocols, which exhibit similarities to \pbft.
This is facilitated by the modular design of the model, \ie, the expression of communication phases, \eg, broadcast, quorum, as \acp{pdf} which can be combined to describe the overall system state.
To demonstrate the adaptability, we show the application of the model to \smart, Zyzzyva and SBFT in Appendices~\ref{app.smart_model_adapt}, \ref{app.zyzzyva_model_adapt}, and~\ref{app.sbft_model_adapt}, respectively.
These adaptations showcase how the model can be applied to a variety of communication patterns,
including client interaction and the possibility to branch into a fast or slow path.

We deliberately chose to highlight the model derivation in this section, leaving the formal definition of the modular components for future work.

\section{Model Validation}\label{sec.validation}
To verify the correctness of the model,
a discrete-event simulator was written in Rust.
The codebase of the simulation is publicly available on Github\footnote{https://github.com/mani2416/bft\_simulation}.
In a first instance, the simulator implements the happy path of \pbft
for single requests without batching.
The dynamic link failure model is realized by discarding each message reception event
with a configurable probability $p_l$ (except the messages from the client). In addition, each node will miss all messages belonging to a certain communication phase with probability $p_c$, simulating a crash failure.
The state transitions of all processes are logged,
\ie, receiving a \texttt{pre-prepare}, collecting a quorum of
\texttt{prepare} and \texttt{commit} messages, as well as any crashes that occurred during each phase.
By doing this, we can compare the simulated state with the predictions of our model.
The simulation can easily scale to larger numbers of nodes (above 100)
since only the state transitions in the happy path are of interest
and no actual \ac{smr} is implemented, \ie, requests are not executed.

In order to validate our model with an independent source,
we also deployed
the Java-based public \ac{bft} \ac{smr} library \smart~\cite{Bessani2014}
as a reference implementation.
Recent work considers the library as a stable implementation of
\ac{bft} \ac{smr}~\cite{Carvalho2018,Rahli2018,Sousa2018}.
It implements a consensus protocol~\cite{Cachin2011a} that bears a high resemblance to \pbft:
it utilizes epochs, an equivalent to the views in \pbft,
and operates in three phases with respective message types.
For simplicity, we stick to \pbft's terminology, when discussing \smart
While mostly similar, the communication pattern of \smart differs from \pbft in two details,
which required minor model adaptations.
Firstly, nodes do not count the primary's \texttt{pre-prepare} message as a \texttt{prepare} message
for the second phase.
Secondly, nodes are allowed to skip the second phase
if a quorum of other nodes was able to complete that phase so that at least $2f+1$ \texttt{commit} messages are broadcast in the third phase.
We describe the model adaptations in Appendix~\ref{app.smart_model_adapt}.

In order to apply dynamic link and crash failures to \smart,
artificial message omission probabilities were implemented into the library.
Accordingly, all messages associated with one of the three phases
are dropped with the probability $p_l$ and each node discards all messages of a whole phase with probability $p_c$.
The \smart implementation uses TCP, though,
and therefore is not directly able to handle dynamic link failures.
As a workaround, we consider each request individually,
simulate the protocol until a failure occurs,
and restart the simulation for every request.
By doing this, we are still able to log all states and state transitions as before.
The changes necessary to implement the aforementioned failures
affected the class that is responsible for handling incoming messages only
and consisted of less than 50 lines of additional code.
The library was executed on a single computer
and up to 10 replicas and one client were instantiated to execute the requests.

Increasing the number of processes while keeping the maximum number of faulty processes constant
leads to an increased robustness of both protocols against link and crash failures,
because more messages are available to build a quorum while the required quorum size remains equal.
We therefore evaluate both protocols for the most interesting scenario $n = 3f+1$,
\ie, the minimum number of processes required to tolerate $f$ faulty processes.

\begin{figure}
	\captionsetup[subfigure]{justification=centering}
	\centering
	\begin{subfigure}{0.33\textwidth}
		\includegraphics[width=\linewidth]{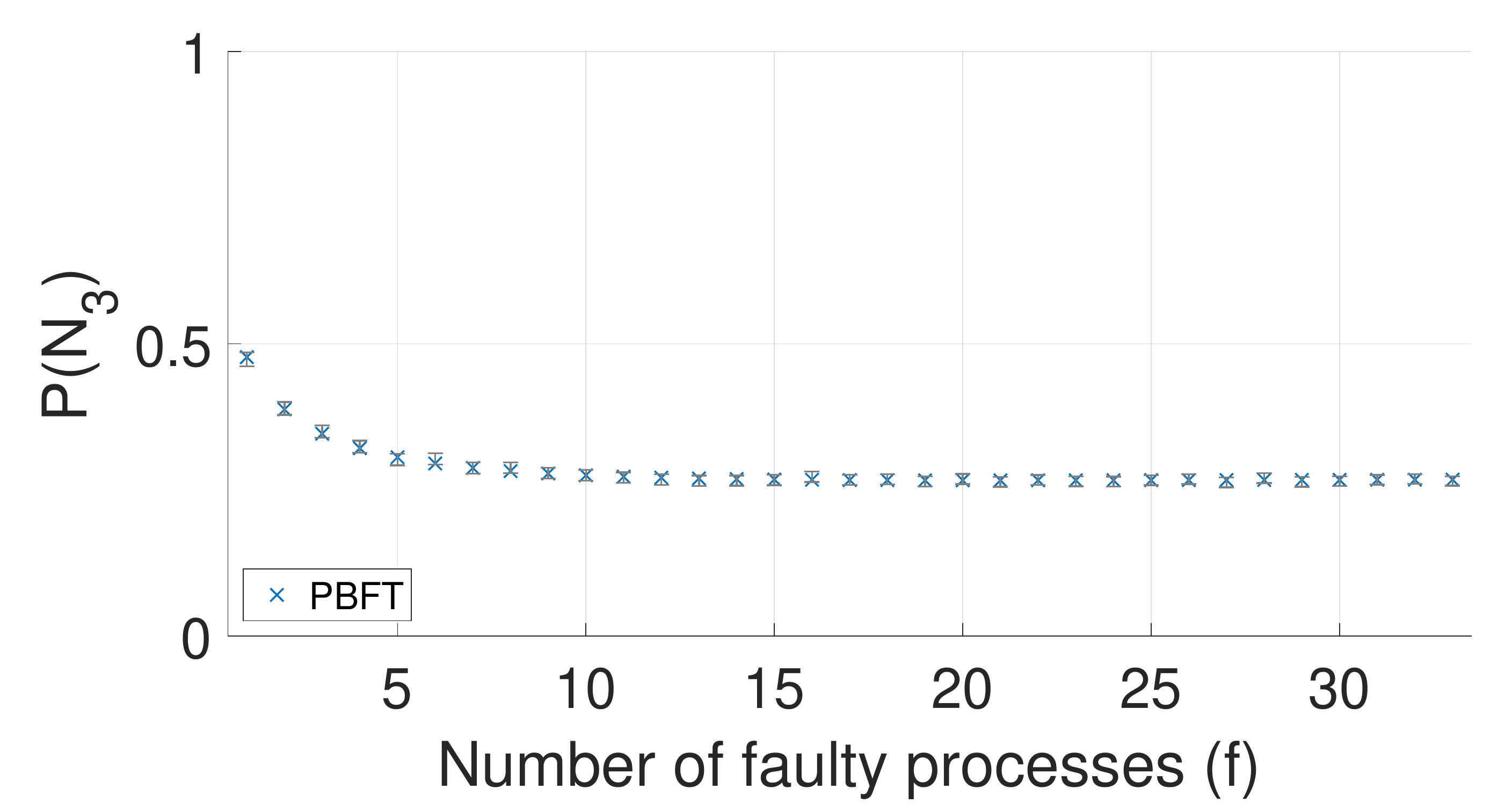}
		\caption{$p_l = p_c = 0.1$}
		\label{fig:pbft_val_f}
	\end{subfigure}\hfil
	\begin{subfigure}{0.33\textwidth}
		\includegraphics[width=\linewidth]{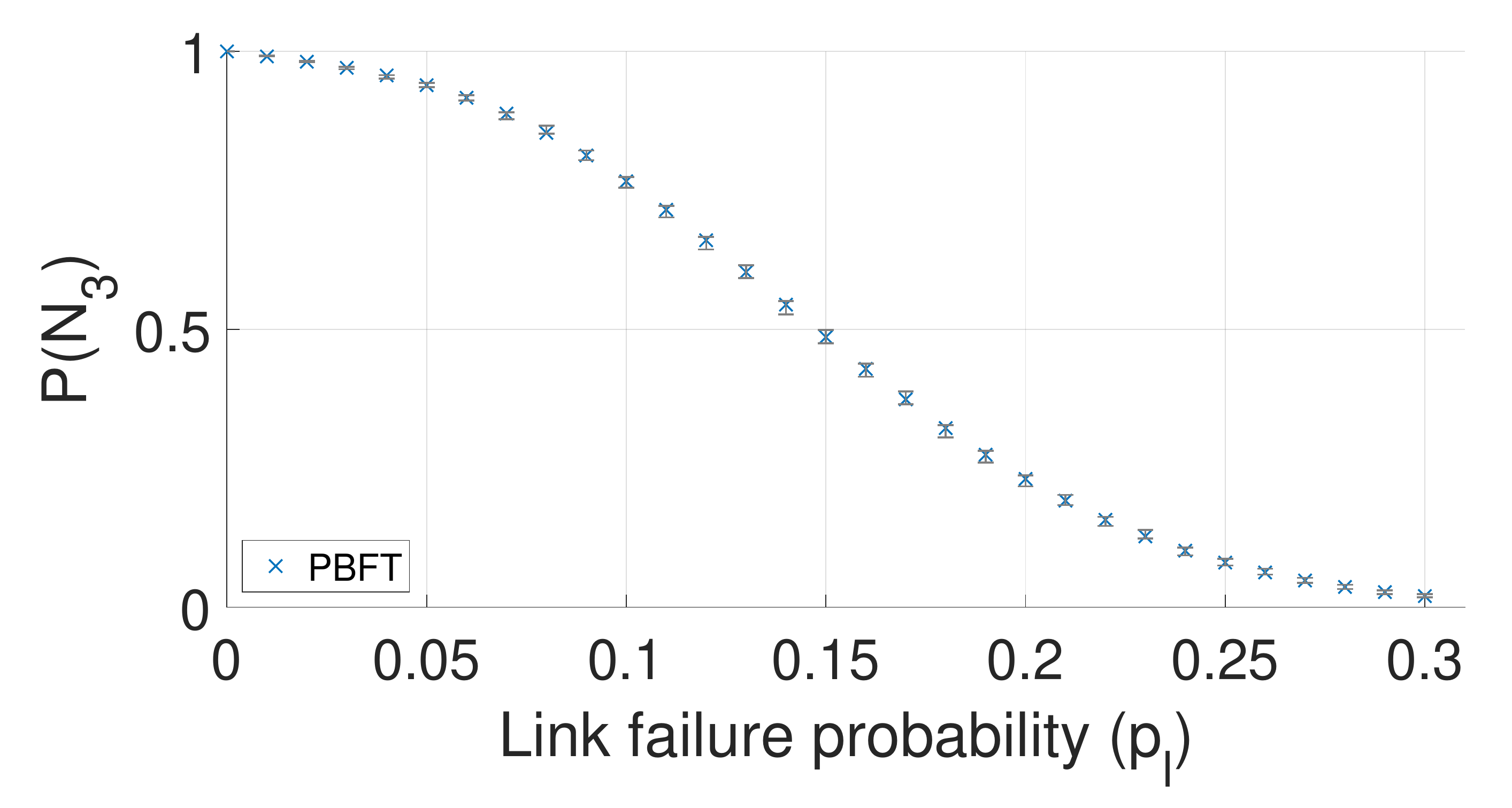}
		\caption{$n = 10$, $p_c = 0$}
		\label{fig:pbft_val_pl}
	\end{subfigure}\hfil
	\begin{subfigure}{0.33\textwidth}
		\includegraphics[width=\linewidth]{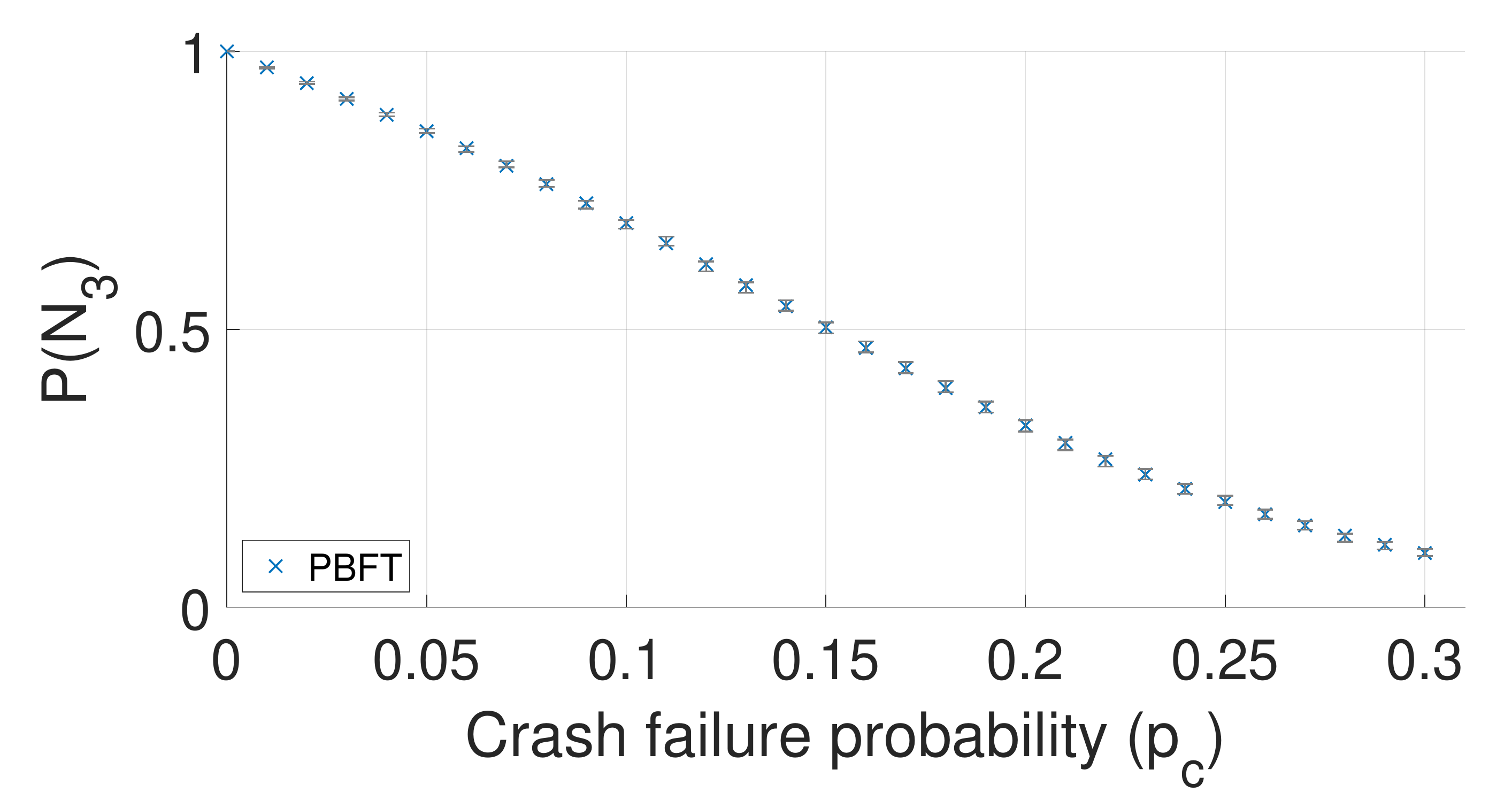}
		\caption{$n = 10$, $p_l = 0$}
		\label{fig:pbft_val_pc}
	\end{subfigure}\hfil
	
	\medskip
	
	\begin{subfigure}{0.33\textwidth}
		\includegraphics[width=\linewidth]{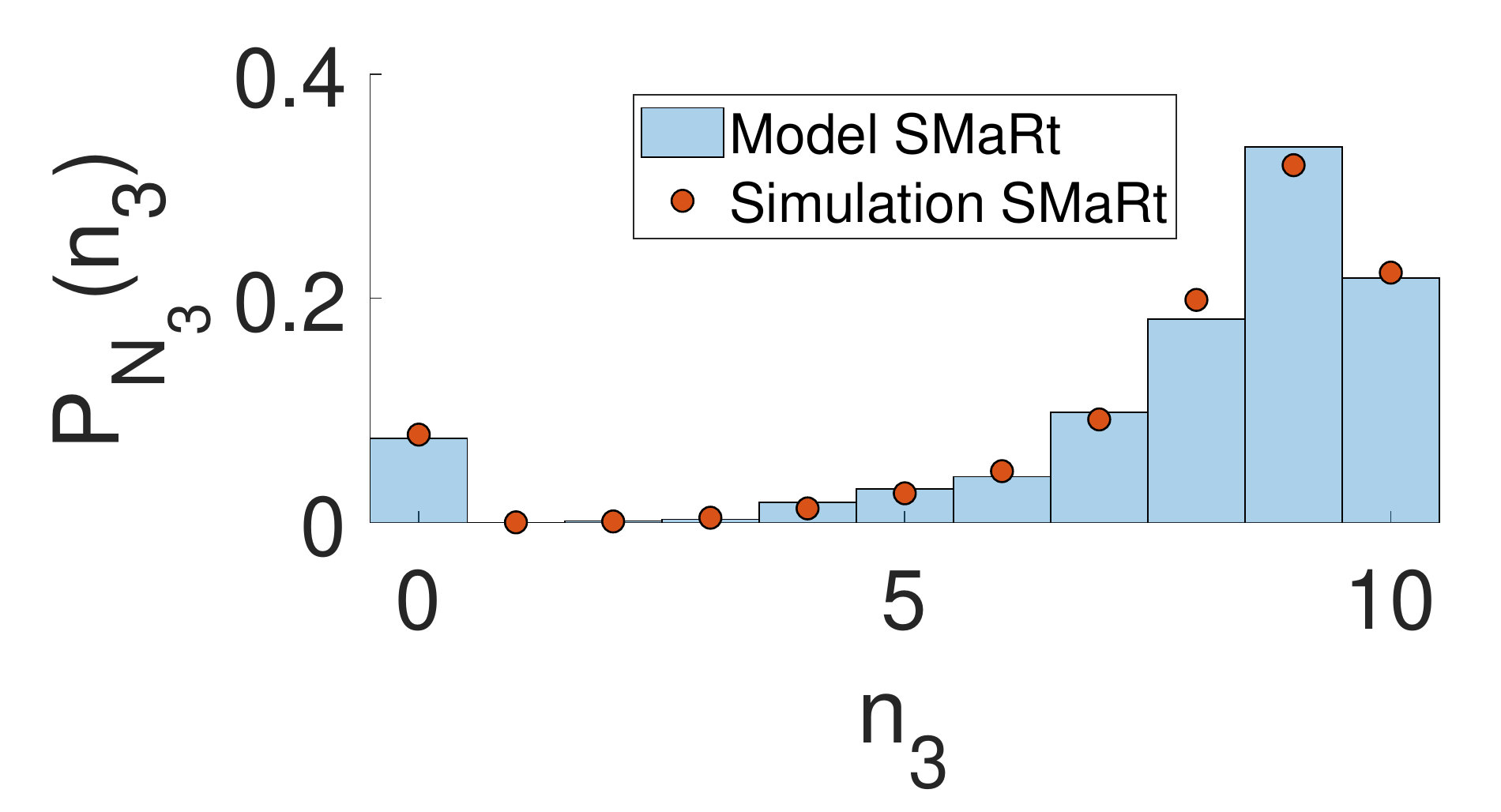}
		\caption{$f = 3, p_l = p_c = 0.05$}
		\label{fig:smart_val_dist}
	\end{subfigure}\hfil
	\begin{subfigure}{0.33\textwidth}
		\includegraphics[width=\linewidth]{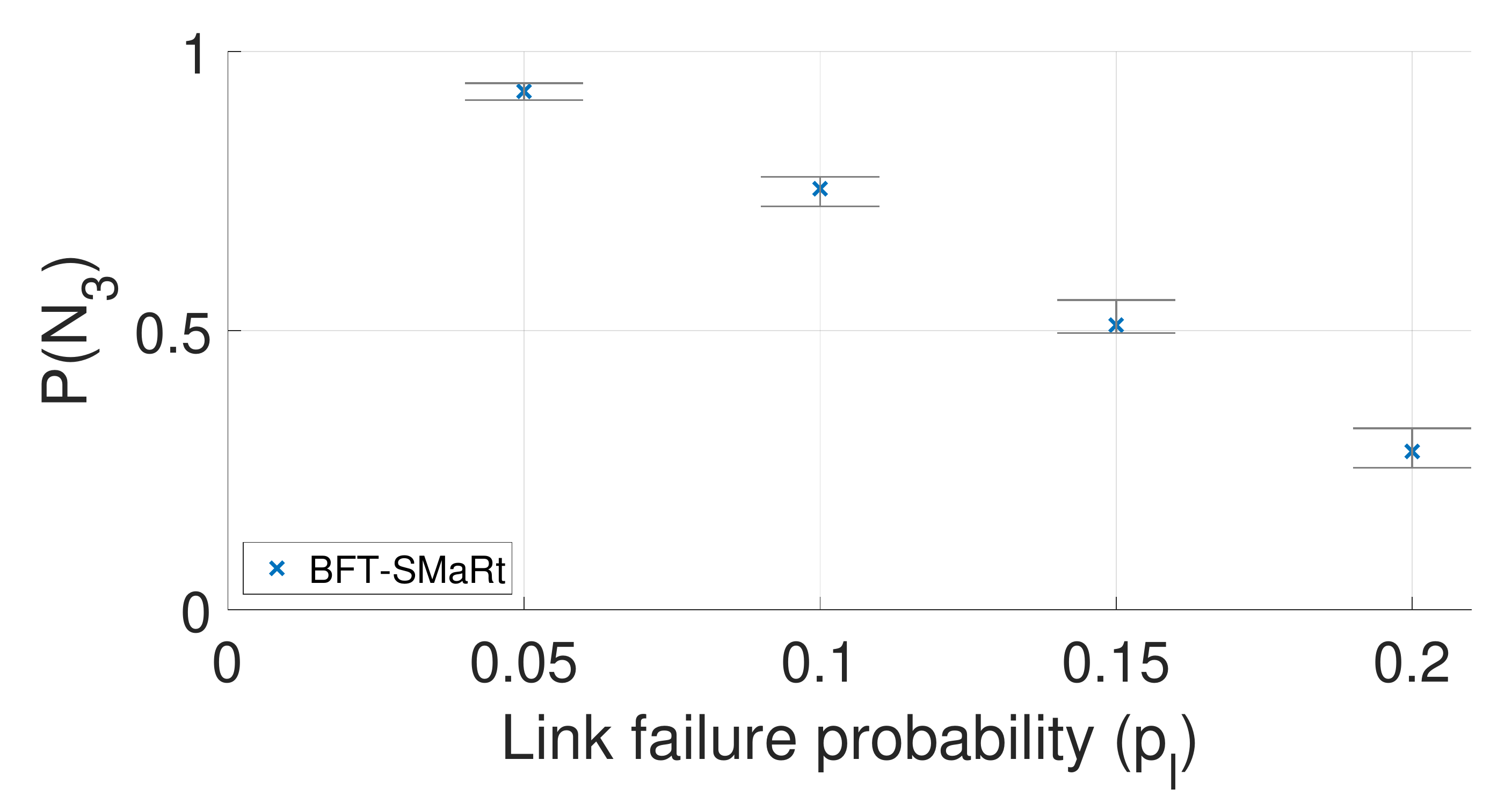}
		\caption{$n = 10$, $p_c = 0$}
		\label{fig:smart_val_pl}
	\end{subfigure}\hfil
	\begin{subfigure}{0.33\textwidth}
		\includegraphics[width=\linewidth]{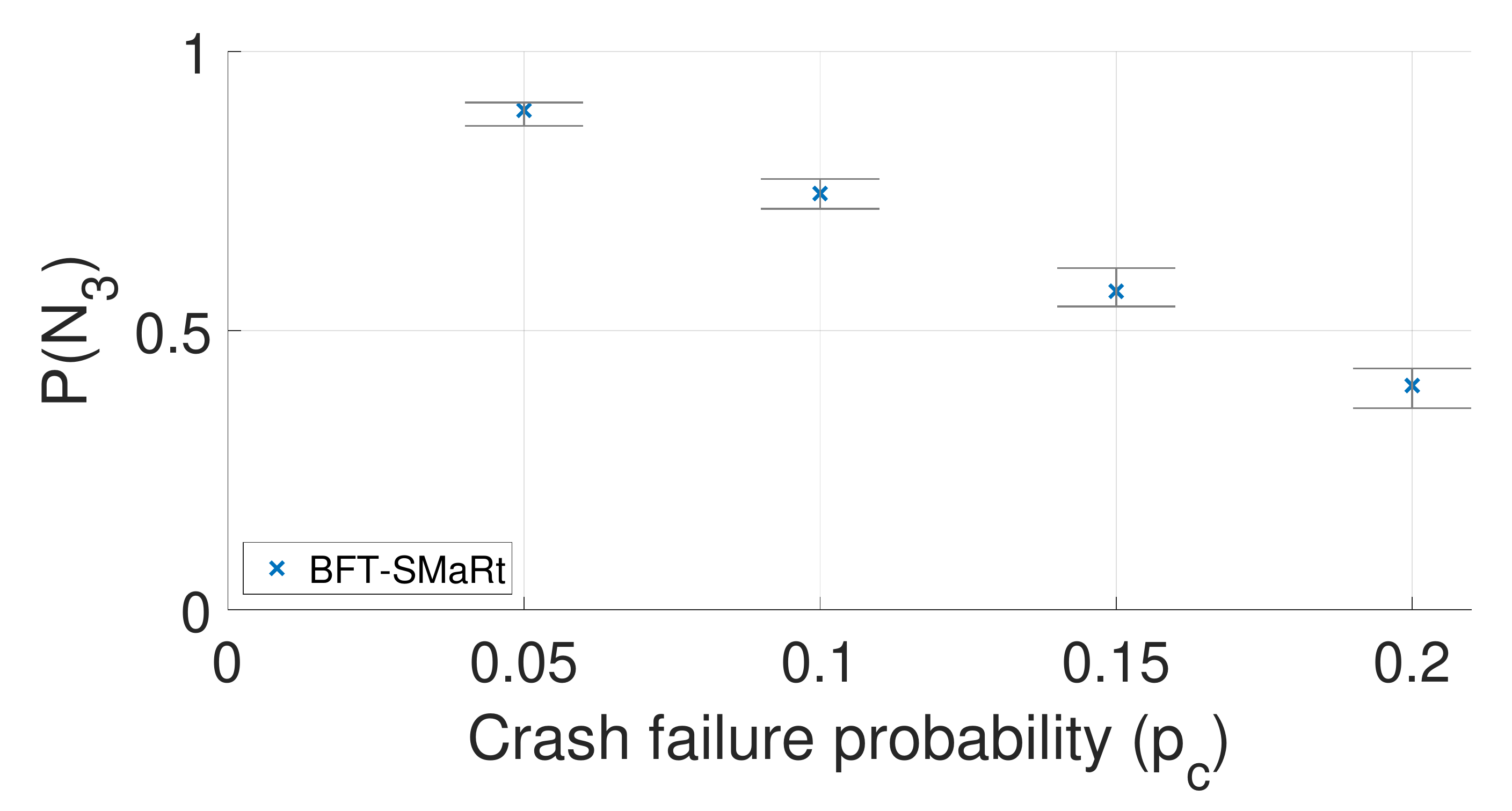}
		\caption{$n = 10$, $p_l = 0$}
		\label{fig:smart_val_pc}
	\end{subfigure}
	\caption{Model validation results for PBFT (a-c) and \smart (d-f).}
	\label{fig.comparison_model_sim_pbft}
\end{figure}

To validate the model for a larger parameter space,
we evaluated \pbft for different numbers of processes, link failure rates, and crash failure rates.
\Cref{fig:pbft_val_f,fig:pbft_val_pl,fig:pbft_val_pc}
show the probability of a single (representative) process
to successfully reach phase $N_3$ for different $n$, $p_l$ and $p_c$, respectively.
The simulation results for 5,000 requests are plotted with $99\%$ confidence intervals and
model predictions are depicted as crosses.
Increasing the number of processes in the network can have, depending on the number of processes and failure rates, either a stabilizing or destabilizing effect on the performance. A more detailed analysis of this behavior is given in \Cref{sub:eval.pbft_smart}.
Increasing the failure rate of either links or processes
causes a constant decrease for $P(N_3)$.

The comparison between model predictions and experimental results for \smart are shown in \Cref{fig:smart_val_dist,fig:smart_val_pl,fig:smart_val_pc}.
Since \smart implements actual \ac{smr} and the execution was unstable due to the previously mentioned halts during the view-change protocol,
we executed 1,000 requests for each parameter combination only (without batching).
\Cref{fig:smart_val_dist} depicts the measured and by the model predicted \ac{pdf} of $P(N_3)$.
The impact of link and crash failures on \smart is similar to \pbft. The small deviations visible between \Cref{fig:pbft_val_pl,fig:pbft_val_pc} and \Cref{fig:smart_val_pl,fig:smart_val_pc} stem from the algorithmic differences described above.
The overall results confirm that our model predictions for \pbft and \smart align accurately with the simulations and experimental results.

\section{Evaluation}\label{sec.evaluation}

\subsection{Protocol stability}
The all-to-all broadcast quorum collection phase in \ac{bft} protocols,
\ie, for ${2f+1}$ processes to collect $2f$ out of $3f$ possible messages (not counting its own),
is inherently resilient against link failures.
A node cannot collect a quorum if at least ${f+1}$ out of its $3f$ incoming links are failing.
Consequently, even in the worst case,
at least ${f+1}$ nodes with at least ${f+1}$ link failures,
\ie ${(f+1)^2}$ overall link failures,
are necessary for the quorum collection phase to potentially fail.

\begin{figure}
	\centering
	\begin{minipage}{.329\textwidth}
		\captionsetup[subfigure]{justification=centering}
		\centering
		\begin{subfigure}{\textwidth}
			\includegraphics[width=\textwidth]{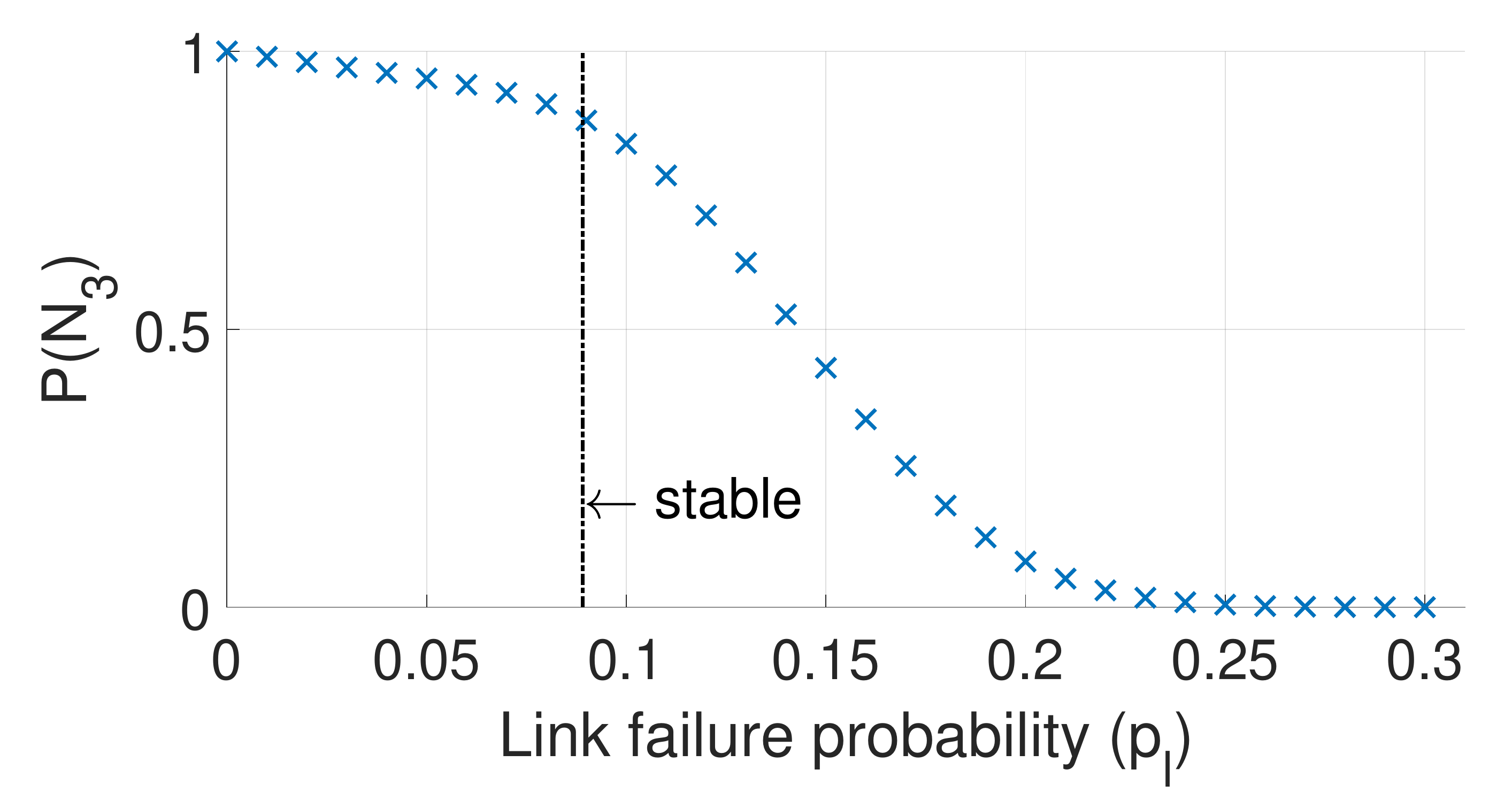}
			\caption{$P(N_3)$ with $p_c = 0$.}
			\label{fig.bound_pl_b}
		\end{subfigure}
		\smallskip
		\begin{subfigure}{\textwidth}
			\includegraphics[width=\textwidth]{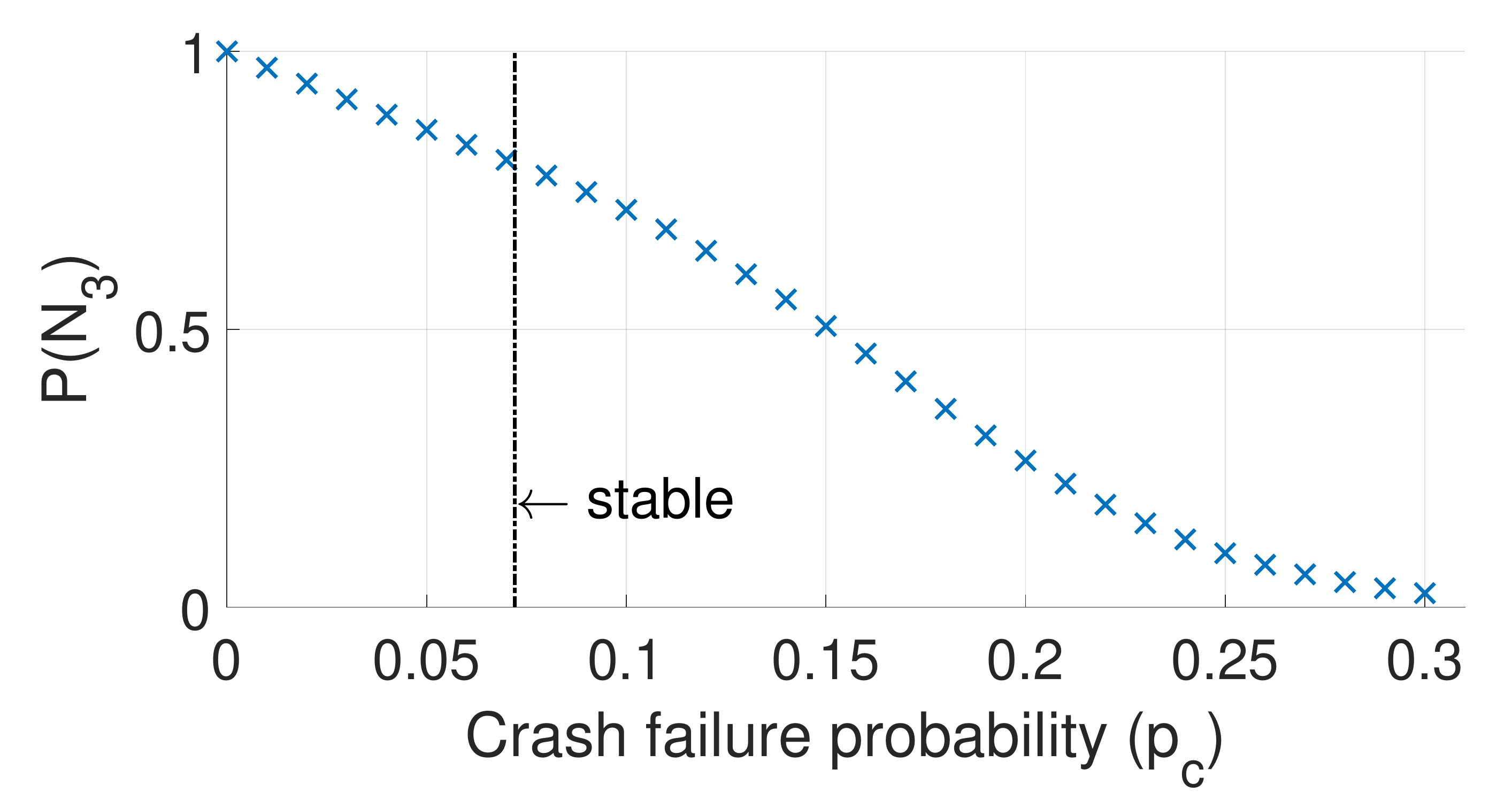}
			\caption{$P(N_3)$ with $p_l = 0$.}
			\label{fig.bound_pc_b}
		\end{subfigure}
		\caption{$P(N_3)$ for \pbft with $n = 25$.}
		\label{fig.boundaries_b}
	\end{minipage}
	\hfill
	\begin{minipage}{.6\textwidth}
		\centering
		\includegraphics[width=\textwidth]{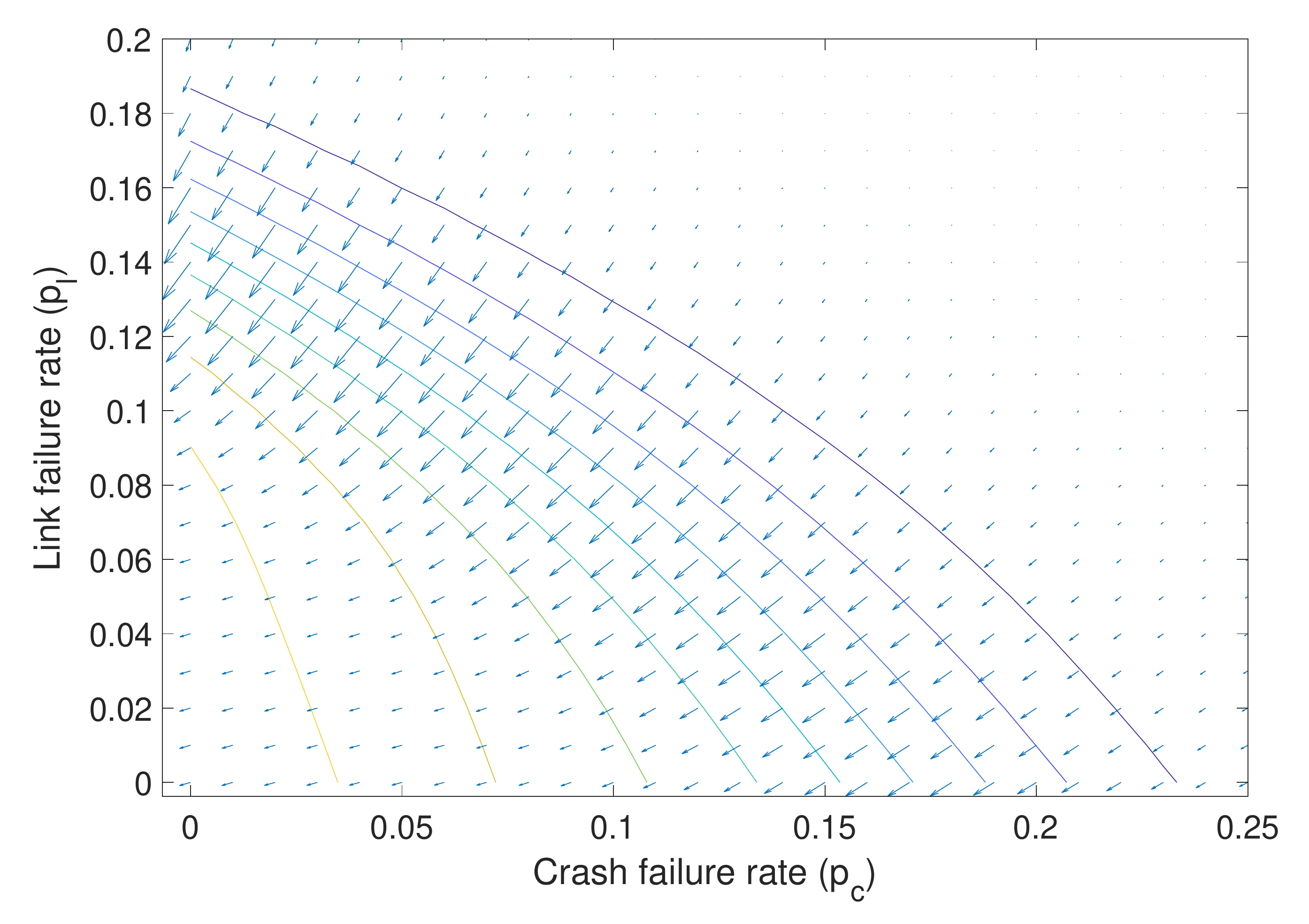}
		\captionsetup{width=\textwidth}
		\caption{Contour plot of $P(N_3)$ as predicted by our model for \pbft with $n=40$; the gradient shows a vector field over $p_c$ and $p_l$.}
		\label{fig.pbft_cont}
	\end{minipage}
\end{figure}

We can validate this theoretical boundary for \pbft with our model.
Since the number of processes that partake in each quorum phase is dependent on previous phases, the boundary for each phase is calculated as
\begin{equation}\label{eq.bound_quorum_stable}
\frac{((f+1)-(n-E[N_{i-1}]))^2}{E[N_{i-1}](E[N_{i-1}]-1)}
\end{equation}
with $E[N_{i-1}]$ being the expected number of nodes that are still active after the previous phase.
Depicted in \Cref{fig.boundaries_b} are the predicted probabilities for $P(N_3)$ for increasing link failures (\Cref{fig.bound_pl_b}) and crash failures (\Cref{fig.bound_pc_b}).
The line labeled "stable" marks the boundary, given by \eqref{eq.bound_quorum_stable}, past which the average number of link failures can potentially cause the quorum phase to fail.
The linear decrease to the left of the boundary originates from the previous phases of the protocol.
In \Cref{fig.bound_pl_b}, the number of processes that receive the \texttt{pre-prepare} message in the first phase ($N_1$) limits the overall number of processes that can complete the protocol.
Since $N_1$ decreases linearly with the link failure rate, so too will $N_2$ and $N_3$.
Past the boundary, the quorum phase may fail due to the increased number of link failures, resulting in an abrupt decline of $P(N_2)$ and $P(N_3)$.
The same behavior is visible for increasing crash failures in \Cref{fig.bound_pc_b}, albeit with an even steeper linear phase.
Because processes cannot recover within the happy path of \pbft, each successive phase with crash failures will decrease the number of available nodes for further phases, leading to the steeper decline before the boundary.

The evaluation methodology and the respective results
can be used to parametrize the protocol to ensure that the
protocol execution remains stable even for a given failure rate.
Since most \ac{bft} protocols treat delayed messages as link failures,
the model can be utilized to fine-tune timeouts.
That is, for a known (or assumed) delay distribution,
a timeout parameter can be translated to a failure rate.
A small timeout leads accordingly to a higher failure rate,
but at the same time is able to quickly detect (genuinely) lost messages
and make progress.
For instance, let us assume that the message delay on all links can be described with a
normal distribution of mean $\mu = \SI{100}{\milli\second}$
and standard deviation $\sigma = \SI{10}{\milli\second}$.
Further, we assume the stable boundary, calculated with \eqref{eq.bound_quorum_stable}, to be $0.1$ for some arbitrary protocol.
The timeout that keeps the protocol in the stable region
is derived by finding an upper bound,
where the integrated \ac{pdf} of the delays is equal to $0.9$.
In our example, the timeout should be $>\SI{87.19}{\milli\second}$.
To conclude, our model allows to evaluate various failure scenarios
and adjust parameters accordingly.

\subsection{Impact of number of processes, link, and node failures in PBFT}
To better demonstrate the predictive capabilities of the model, a contour plot of $P(N_3)$ for \pbft is provided in \Cref{fig.pbft_cont}, for varying link and crash failure rates.
Additionally, the gradient is displayed, derived from the operating points for different $p_l$ and $p_c$, as they are predicted by the model.
The orientation of the arrows indicates the impact of variations in either failure rate on $P(N_3)$.
The more pronounced the horizontal component of a vector, the more dominant is the impact of crash failures on $P(N_3)$ and the same applies to the vertical component and link failures.
The contour plot allows to quickly discern the impact of either failure rate on the protocol performance.
For example, for low link failure rates, \ie, $p_l < 0.05$, changes in the link failure rate have mostly a negligible impact compared to changes in the crash failure rate.
However, for very low rates of crash failures and a moderate number of link failures ($p_l > 0.1$), the link failure rate dominates the performance of \pbft.

\Cref{fig.pbft_cont} also validates the observations made in \Cref{fig.bound_pl_b}.
While the link failure rate is below the boundary, \ie, the linear slope, the crash failure rate mostly dominates the impact on the performance.
For higher values of link failures, their impact increases, correlating to the steeper decline that follows after the boundary.

The influence of process numbers for a given link failure rate on \ac{pbft} is plotted in \Cref{fig:eval_pbft_dip}.
While it is well known that most \ac{bft} protocols do not scale well with the number of processes
due to the quadratic message complexity,
it generally offers means to increase stability in the presence of dynamic link failures.
In Appendix~\ref{app.proof_conv_binomial}, we show that
the probability to collect a quorum for $n \rightarrow \infty$ converges to $0$ or $1$,
depending on $p_l$ and the quorum size.
We also observe a dip in $P(N_3)$,
visible for fewer nodes and failure rates between $9\%$--$15\%$.
The dip is caused by two effects:
(i)~for larger $n$ the variance of the binomial distribution increases, and
(ii)~while the mean of the distribution narrows down on a value depending on the failure rates,
the number of required messages to complete the quorum moves towards a discrete value,
which is approximately $\frac{2}{3}$.
As a consequence, the number of nodes can increase the success probability of the quorum collection phase for failure rates below a certain threshold.

\subsection{Comparison between PBFT, BFT-SmaRt, Zyzzyva, and SBFT}\label{sub:eval.pbft_smart}

\begin{figure}[t]
	\captionsetup[subfigure]{justification=centering}
	\centering
	\begin{subfigure}{0.5\textwidth}
		\includegraphics[width=\linewidth]{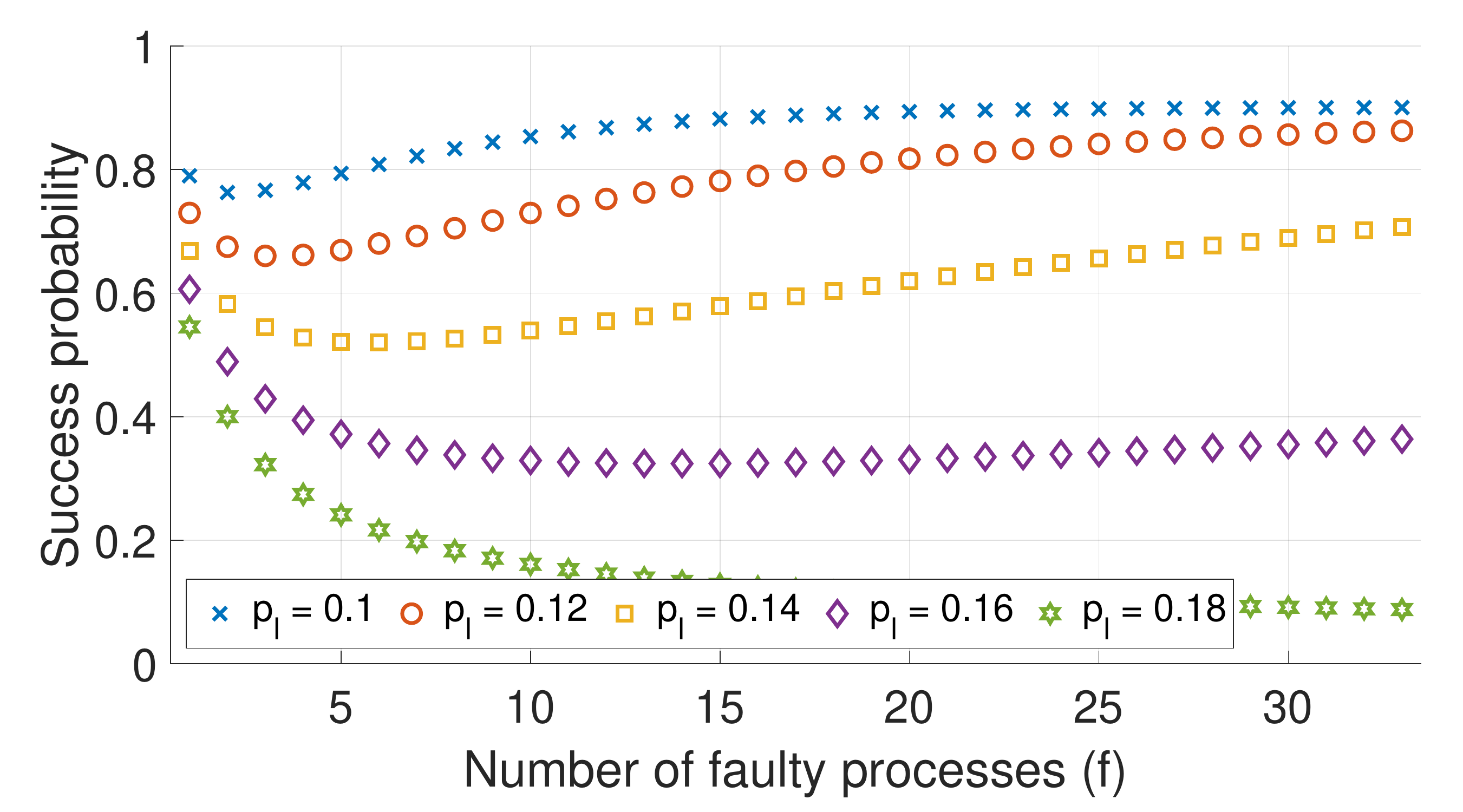}
		\caption{$p_c = 0.$}
		\label{fig:eval_pbft_dip}
	\end{subfigure}\hfil
	\begin{subfigure}{0.5\textwidth}
		\includegraphics[width=\linewidth]{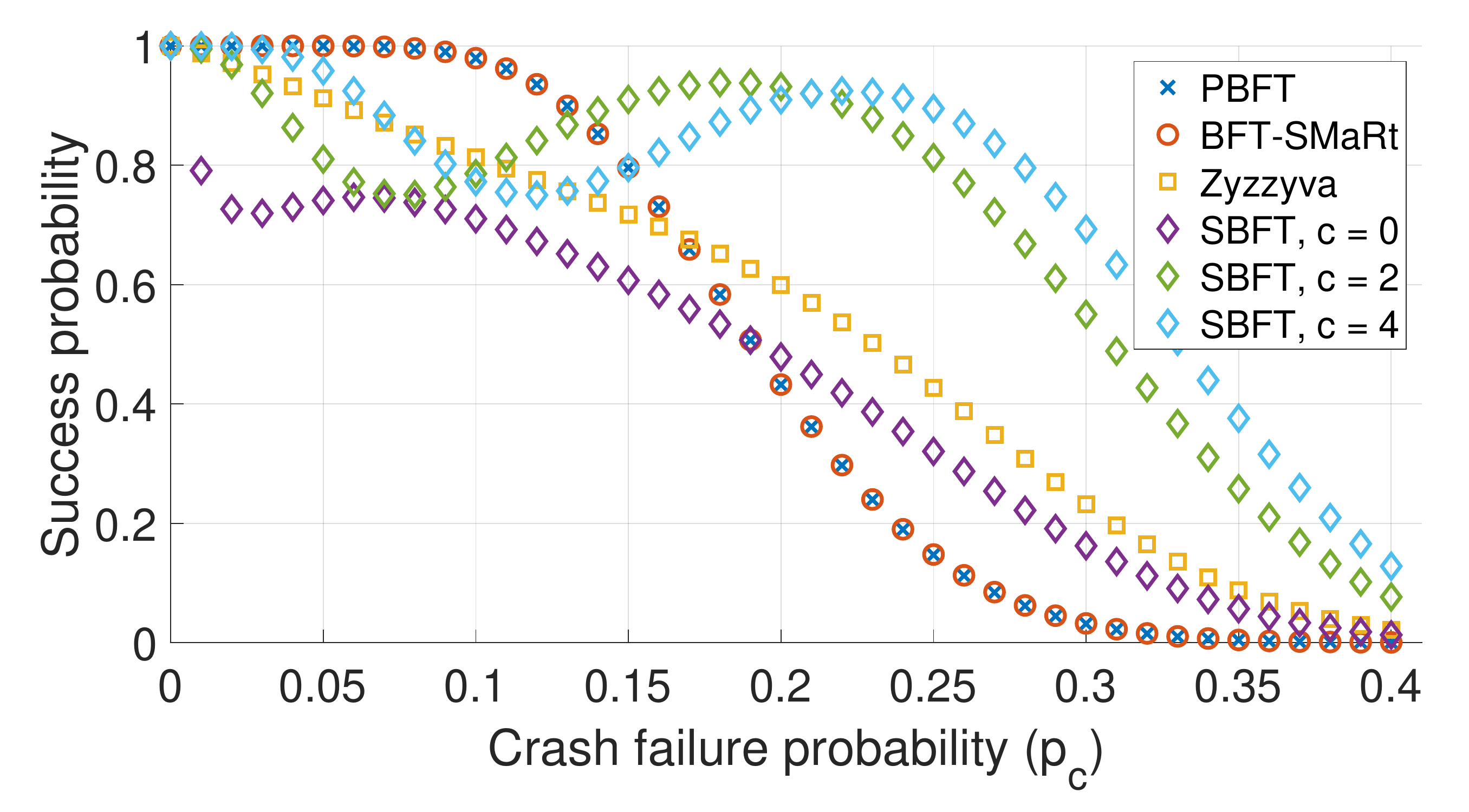}
		\caption{$f = 10, p_l = 0.$}
		\label{fig:eval_all_crash}
	\end{subfigure}

	\medskip

	\begin{subfigure}{0.5\textwidth}
		\includegraphics[width=\linewidth]{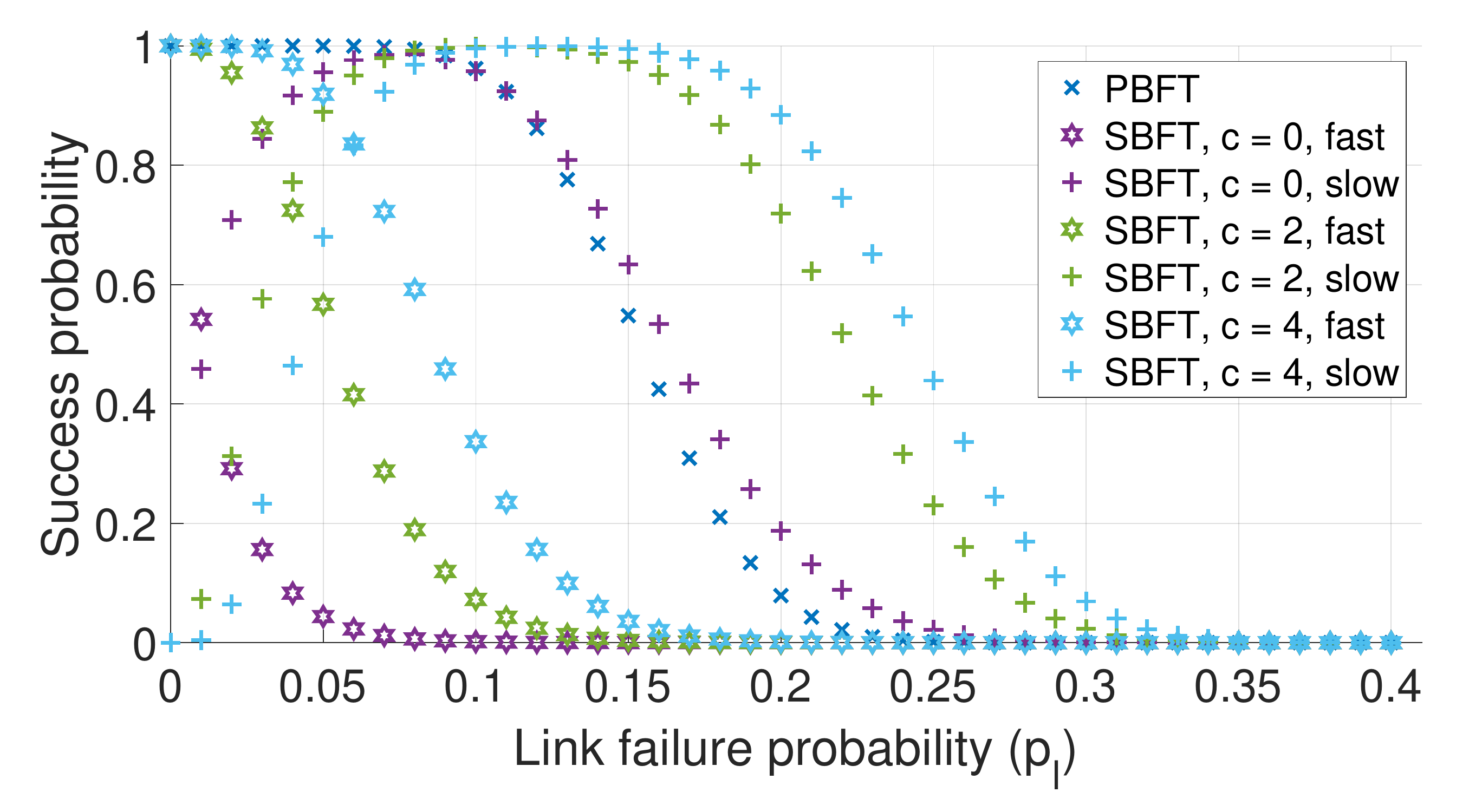}
		\caption{$f = 10, p_c = 0.$}
		\label{fig:eval_sbft_link}
	\end{subfigure}\hfil
	\begin{subfigure}{0.5\textwidth}
		\includegraphics[width=\linewidth]{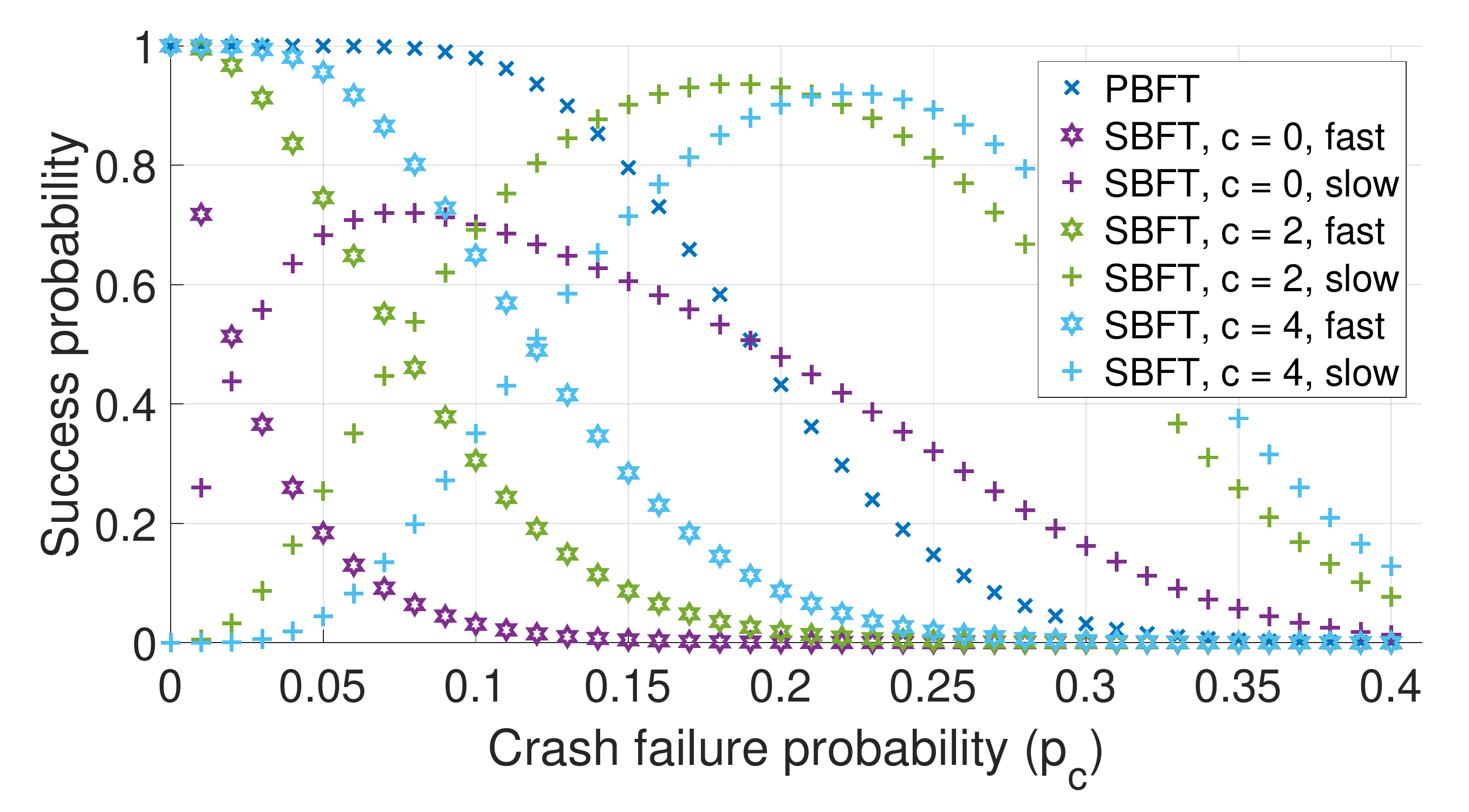}
		\caption{$f = 10, p_l = 0.$}
		\label{fig:eval_sbft_crash}
	\end{subfigure}

	\caption{Predictions made by our model: (a) depicts the impact of number of replicas for different link failure rates in PBFT, (b) compares the happy path success probabilities of PBFT, BFT-SMaRt, Zyzzyva as well as SBFT and a more detailed analysis of SBFT is given with (c) and (d).}
	\label{fig:evaluation_protocols}
\end{figure}

To showcase the adaptability of our model, we applied it to Zyzzyva~\cite{Kotla2007} and SBFT~\cite{GolanGueta2019}.
The detailed adaptations to the model are listed in \Cref{app.zyzzyva_model_adapt,app.sbft_model_adapt}.

An exemplary comparison of all protocols for different crash failure rates
is given in \Cref{fig:eval_all_crash}.
Depicted are the overall success rates,
\ie, for the Zyzzyva and SBFT the combination of fast and slow paths.
To better demonstrate the capabilities of the model,
the individual success probabilities for the fast and slow path of SBFT for different link failure rates are plotted in \Cref{fig:eval_sbft_link,fig:eval_sbft_crash}.
Since SBFT allows for optional, additional replicas,
denoted as $c$,
the model allows to quickly assess the protocol behavior for different failure rates and configurations of $c$.
The plots show that SBFT outperforms PBFT for increasing numbers of $c$ and higher failure rates,
while PBFT is more stable if SBFT transitions from the fast path to the slow path.

\section{Conclusion}\label{sec.summary}
The probabilistic predictions of the presented model were validated with implementations of \pbft and \smart for various numbers of processes and dynamic link and crash failure rates.
It was demonstrated with \smart, Zyzzyva and SBFT, that the model can be adapted with little effort to other communication patterns of BFT protocols.
The model gives a prediction of the distribution of process states during execution, allowing prediction of protocol behavior (e.g. how many view changes will occur) and therefore performance evaluation.
Additionally, if the  message delay statistics are known, the model can be deployed to tune the timeouts for \ac{bft} protocols, since most protocols cannot differentiate between a delayed or an omitted message, making them indifferent in their impact on the performance of the algorithm.
The model allows to assess the impact of crash and link failures for various operating points of a protocol to identify key boundaries regarding protocol stability.

As was demonstrated with \smart, Zyzzyva and SBFT, the model can be applied to different \ac{bft} protocols by modifying the respective equations for the distributions or adding further random variables should the protocol consist of more phases (as is the case with SBFT).
Further adaptations are facilitated by the fact that a body of \ac{bft} protocols are derived from the core structure of \pbft and consist of interdependent phases.

In further work, we are planning to apply the model to more BFT protocols and evaluate their performance regarding dynamic failures.
Furthermore we are exploring possibilities to extend the model to predict
more sophisticated key performance indicators, such as throughput and latency.
Lastly, we will consider adaptations to our model in order to account for correlated link failures, \eg, as was proposed with a model by Nguyen~\cite{Nguyen2019}.

\clearpage
\bibliography{../../bibliography/consensus}

\clearpage
\appendix
\section{Model adaptations for \smart}
\label{app.smart_model_adapt}

In order to accurately capture the communication pattern of \smart, two adaptations to the model, compared to \pbft, are necessary.
Firstly, the collection of \texttt{prepare} messages is not optimized
to count the primary's \texttt{pre-prepare} as a \texttt{prepare} message.
Secondly, in \pbft, the processes can only successfully complete a consensus round,
if they have reached all three stages of the protocol,
\ie, received a \texttt{pre-prepare} message and collected $2f+1$ of both, \texttt{prepare} and \texttt{commit} messages.
Complementary to this, in \smart,
a process can complete a consensus round,
if it has received a \texttt{pre-prepare} message and collected $2f+1$ \texttt{commit} messages.
Since the collection of $2f+1$ \texttt{commit} messages implies
that at least $f+1$ processes have received $2f+1$ \texttt{prepare} messages,
the processes can commit safely.

Applying these differences to the model results in changes to two equations:
the calculation of $C_2$ in \eqref{eq.c2_conditional} and $C_3$ in \eqref{eq.c3_conditional}.\\
\textbf{Second phase:} The replicas send out \texttt{prepare} messages.
\begin{align}
\label{eq.c2_conditional_smart}
P_{C_2|N_1}(c_2 \con n_1) = B(n_1+1,p_2(n_1),c_2)\\
\label{eq.c2_conditional_helper_smart}
p_2(n_1) = B(n_1,1-p_l,[2f, n]).
\end{align}
\textbf{Third phase:} The replicas send out \texttt{commit} messages.
\begin{align}\label{eq.c3_conditional_smart}
&P_{C_3|N_1, N_2}(c_3 \con n_1, n_2) = \sum\limits_{a+b = c_3} B(n_2, p_a(n_2), a) \cdot B(n_1+1-n_2, p_b(n_2), b)\\
&p_a(n_2) = B(n_2-1, 1-p_l, [2f, n_2-1])\\
&p_b(n_2) = B(n_2, 1-p_l, [2f+1, n_2])
\end{align}
As \eqref{eq.c3_conditional_smart} is now conditioned on two previous phases, the calculation of the distribution has to be adapted with
\begin{align}
&P_{C_3}(c_3) = \sum\limits_{n_1}\sum\limits_{n_2} P(C_3 = c_3 \con N_1=n_1, N_2=n_2) \cdot P_{N_2|N_1}(n_2 \con n_1) \cdot P_{N_1}(n_1)\\
\label{eq.smart_n2n1}
&P_{N_2|N_1}(n_2 \con n_1) = \sum\limits_{c_2} P_{N_2|C_2}(n_2 \con c_2) P_{C_2|N_1}(c_2 \con n_1).
\end{align}

\section{Model adaptations for Zyzzyva}
\label{app.zyzzyva_model_adapt}
Zyzzyva\cite{Kotla2007} uses a primary to order requests,
afterwards, the client communicates directly with the replicas to collect either $3f+1$ (fast path) or two quorums of $2f+1$ (slow path) messages.
Due to the direct integration of the client into the communication patterns,
the link and crash failure rates are assumed to apply to the client as well.
The impact of crash failures, i.e. the calculation $N_i \rightarrow C_i$, is identical for all phases.\\
\textbf{First phase:} the primary broadcasts a \texttt{prepare} to all replicas, identical to \ac{pbft}.\\
\textbf{Second phase:} the replicas respond  to the client (quorum of $3f+1$ for fast path or at least $2f+1$ for slow path).
\begin{align}
&P_{C_2|N_1, \text{fast}}(c_2|n_1) = B(1, p_{2,f}(n_1), c_2)\\
&p_{2,f}(n_1) = B(n_1, 1-p_l, 3f+1)\\
&P_{C_2|N_1, \text{slow}}(c_2|n_1) = B(1, p_{2,s}(n_1), c_2)\\
&p_{2,s}(n_1) = B(n_1, 1-p_l, [2f+1, 3f])
\end{align}
\textbf{Third phase:} the client broadcasts to all replicas.
\begin{align}
&P_{C_3|N_2}(c_3|n_2) = B(n, p_3(n_2), c_3)\\
&p_3(n_2) = B(n_2, 1-p_l, 1)
\end{align}
\textbf{Fourth phase:} the replicas respond to the client (quorum of $2f+1$).
\begin{align}
&P_{C_4|N_3}(c_4|n_3) = B(1, p_4(n_3), c_4)\\
&p_4(n_3) = B(n_3, 1-p_l, [2f+1, n])
\end{align}
The success probabilities for the fast and slow path are $P(C_2 = 1)$ and $P(C_4 = 1)$, respectively.

\section{Model adaptations for SBFT}
\label{app.sbft_model_adapt}
SBFT \cite{GolanGueta2019} combines the properties of optimistic execution, i.e. all replicas execute in a fast path,
and redundancy, i.e. additional, optional replicas.
In SBFT, the network consists of $3f+2c+1$ replicas, with $c$ being redundant replicas. The algorithm utilizes threshold signatures to reduce the message complexity for collecting the required quorum sizes ($3f+c+1$ for the fast path and $2f+c+1$ for the slow path).
SBFT uses a primary to order requests.
The impact of crash failures, i.e. the calculation $N_i \rightarrow C_i$, is identical for all phases.\\
\textbf{First phase:} the primary broadcasts a \texttt{prepare} to all replicas, identical to \ac{pbft}.\\
\textbf{Second phase:} the replicas respond to the collectors.
\begin{align}
&P_{C_2|N_1, \text{fast}}(c_2|n_1) = B(c+1, p_{2,f}(n_1), c_2)\\
&p_{2,f}(n_1) = B(n_1, 1-p_l, [3f+c+1, n_1])\\
&P_{C_2|N_1, \text{slow}}(c_2|n_1) = B(c+1, p_{2,s}(n_1), c_2)\\
&p_{2,s}(n_1) = B(n_1, 1-p_l, [2f+c+1, 3f+c])
\end{align}
\textbf{Third phase:} the collectors broadcast to all replicas (fast and slow path identical).
\begin{align}
&P_{C_3|N_2}(c_3|n_2) = \sum\limits_{n_2+k = c_3} n_2 + B(n-n_2, p_2(n_2), k)\\
&p_2(n_2) = B(n_2, 1-p_l, [1, n_2])
\end{align}
\textbf{Fourth phase:} the replicas respond to the collectors (fast path end).
\begin{align}
&P_{C_4|N_3, \text{fast}}(c_4|n_3) = B(c+1, p_{4,f}(n_3), c_4)\\
&p_{4,f}(n_3) = B(n_3, 1-p_l, [f+1, n_3])\\
&P_{C_4|N_3, \text{slow}}(c_4|n_3) = B(c+1, p_{4,s}(n_3), c_4)\\
&p_{4,s}(n_3) = B(n_3, 1-p_l, [2f+c+1, n_3])
\end{align}
\textbf{Fifth phase:} the collectors broadcast to all replicas.
\begin{align}
&P_{C_5|N_4, N_1}(c_5|n_4, n_1) = \sum\limits_{n_4+k = c_5} n_4 + B(n_1-n_4, p_4(n_4), k)\\
&p_4(n_4) = B(n_4, 1-p_l, [1, n_4])
\end{align}
the conditional probability $P_{N_4|N_1}(n_4|n_1)$ is calculated in the same manner as \eqref{eq.smart_n2n1}.\\
\textbf{Sixth phase:}  the replicas respond to the collectors (slow path end).
\begin{align}
&P_{C_6|N_5}(c_6|n_5) = B(c+1, p_6(n_5), c_6)\\
&p_6(n_5) = B(n_5-1, 1-p_l, [f, n_5-1])
\end{align}

\section{Convergence of 'binomial quorums'}
\label{app.proof_conv_binomial}

The probability of a replica to collect a quorum of $k$ messages, if each of the $n$ incoming messages has an independent probability $p$ to be omitted, can be formulated as
\begin{align}
P(Q > k) =\sum\limits_{i=k}^{n} B(n,1-p,i),
\end{align}
which, considering the complementary event, can be rewritten as
\begin{align}\label{eq.quorum_base}
P(Q > k) = \sum\limits_{i=0}^{n-k} B(n,p,i).
\end{align}
The binomial distribution can be approximated with the central limit theorem, or in this case, the \texttt{De Moivre-Laplace theorem}.
The theorem states, that a binomial distribution $B(n,p)$ for $p \neq {0, 1}$ will converge to a normal distribution as $n$ grows to infinity.
The normal distribution has a mean of $\mu = np$ and standard deviation of $\sigma = \sqrt{np(1-p)}$.\\
The binomial distribution can thus be approximated as
\begin{align}\label{eq.normal_distribution_approx}
\begin{split}
B(n,p) &= \binom{n}{k} \cdot p^k (1-p)^k\\
&= \frac{1}{2\pi p(1-p)}\cdot \exp{\left(- \frac{(k-np)^2}{2np(1-p)}\right)}
\end{split}
\end{align}
This leads to a cumulative distribution function of
\begin{align}\label{eq.normal_distribution_cumulative}
F(x) = \frac{1}{2}\left[ 1 + \erf\left( \frac{x-np}{\sqrt{2\cdot np (1-p)}} \right) \right]
\end{align}
Applying this formula to (\eqref{eq.quorum_base}) leads to
\begin{align}
\begin{split}
P(Q > k) &= F(x=(n-k))\\
&= \frac{1}{2}\left[ 1 + \erf\left( \frac{(n-k)-np}{\sqrt{2\cdot np(1-p)}} \right) \right]
\end{split}.
\end{align}
If $n$ grows to infinity, we get
\begin{align}
\begin{split}
\lim_{n \rightarrow \infty} P(Q > k) = \frac{1}{2}\left[ 1 + \erf\left( \frac{n-k-np}{\sqrt{2\cdot np(1-p)}} \right) \right]
\end{split}.
\end{align}
Of interest here are the development of the numerator and denominator in the exponential term.
The behavior of the numerator for growing $n$ depends on the value of $p$ and $k$.
Considering that $k$, the size of the quorum, has to be a value between 1 and $n$, we can substitute $k$ with $\lceil qn \rceil$, with $q$ being the relative size of the quorum compared to $n$.
\begin{align}
\begin{split}
\lim_{n \rightarrow \infty} P(Q > k) = \frac{1}{2}\left[ 1 + \erf\left( \frac{n(1-q-p)}{\sqrt{2\cdot np(1-p)}} \right) \right]
\end{split}.
\end{align}
For the nominator applies
\begin{align}
\lim_{n \rightarrow \infty} n(1-q-p) = \begin{cases}
- \infty &, \text{ for } p > 1-q\\
\infty &, \text{ for } p < 1-q\\
0 &, \text{ for } p = 1-q
\end{cases}.
\end{align}
Meanwhile, the denominator will always converge towards infinity, albeit slower than the numerator
\begin{align}
\lim_{n \rightarrow \infty} \sqrt{2\cdot np(1-p)} \rightarrow \sqrt{\infty}
\end{align}
Since the error function has the properties
\begin{align}
&\lim_{x\rightarrow \infty} \erf(x) = 1,\\
&\lim_{x\rightarrow -\infty} \erf(x) = -1,\\
&\erf(0) = 0,
\end{align}
we can conclude that
\begin{align}\label{eq.quorum_lim_final}
\begin{split}
&\lim_{n \rightarrow \infty} P(Q > k) =\\
&\begin{cases}
\frac{1}{2}\left[ 1 + \erf(-\infty) \right] = 0, &\text{ for } p > 1-q\\
\frac{1}{2}\left[ 1 + \erf(\infty) \right] = 1, &\text{ for } p < 1-q\\
\frac{1}{2}\left[ 1 + \erf(0) \right] = \frac{1}{2}, &\text{ for } p = 1-q
\end{cases}
\end{split}
\end{align}
with $q \approx \frac{2}{3}$ for \ac{bft}.

\end{document}